\documentstyle[12pt]{article}
\newcommand{\nc}{\newcommand}
\nc{\beq}{\begin{equation}}
\nc{\eeq}{\end{equation}}
\nc{\beqa}{\begin{eqnarray}}
\nc{\eeqa}{\end{eqnarray}}
\nc{\lra}{\leftrightarrow}
\def\sfrac#1#2{{\textstyle\frac{#1}{#2}}}
\def\sqrtwo{{\scriptstyle\sqrt{2}\,}}
\nc{\sss}{\scriptscriptstyle}
{\nc{\lsim}{\mbox{\raisebox{-.6ex}{~$\stackrel{<}{\sss\sim}$~}}}
{\nc{\gsim}{\mbox{\raisebox{-.6ex}{~$\stackrel{>}{\sss\sim}$~}}}

\def\ppp{\phantom{+}}

\begin{document}
\input epsf.tex
\title{\vskip2cm
{\bf Was the Electroweak Phase Transition \\ Preceded by
a Color-Broken Phase?}}

\author{ James~M.~Cline\footnote{e-mail:
        jcline@physics.mcgill.ca} ${}^{,a}$,
Guy D.~Moore\footnote{e-mail: guymoore@physics.mcgill.ca} ${}^{,a}$, and
G\'eraldine Servant\footnote{e-mail: servant@physics.mcgill.ca} ${}^{,a,b}$
\hspace{3cm}\\
${}^a$ {\small Dept.~of Physics, McGill University, 3600 University St.}\\
        {\small Montreal, Qc H3A 2T8 Canada}\\
${}^b$ {\small Service de physique th\'eorique du CEA Saclay}\\
        {\small 91191 Gif sur Yvette c\'edex, France}\\}

\maketitle \begin{abstract} It has been suggested, in connection with
electroweak baryogenesis in the Minimal Supersymmetric Standard Model
(MSSM), that the right-handed top squark has a negative mass squared
parameter, such that its field could condense prior to the electroweak
phase transition (EWPT).  Thus color and electric charge could have
been broken just before the EWPT.  Here we investigate whether the
tunneling rate from the color-broken vacuum can ever be large enough
for the EWPT to occur in this case.  We find that, even when all
parameters are adjusted to their most favorable values, the nucleation
rate is many orders of magnitude too small.  We conclude that, without
additional physics beyond the MSSM, the answer to our title question is
``no.''  This gives constraints in the plane of the light stop mass
versus parameters related to stop mixing.  However it may be possible
to get color breaking in extended models, such as those with $R$-parity
violation.  \end{abstract}

\vskip1cm
\leftline{}
\leftline{}

\vskip-20cm
\rightline{}  
\rightline{MCGILL-99/01}
\rightline{hep-ph/9902220}

\newpage

\section{Introduction}

The baryon asymmetry of the universe (the excess of baryons over
antibaryons) is a very interesting puzzle, and it is exciting that its
resolution may involve only electroweak physics which is either known or
testable at colliders in the near future.
This is because electroweak physics has the potential for  satisfying
all three of Sakharov's conditions \cite{Sakharov} for baryogenesis.
The first, baryon number nonconservation, occurs because of the anomaly
and the topological structure of the vacuum in the SU(2) sector of the
electroweak theory \cite{tHooft}; further, baryon number violation
becomes quite efficient at high temperatures, $T > 100$  GeV
\cite{bviol}.  The second condition, CP violation, is present
but insufficient in the minimal standard model \cite{Gavela};
however, there are new sources in some extended models which allow for
enough baryon production.  

The third condition is that
baryon number violating processes are out of thermal equilibrium, at
the moment of baryogenesis.  Electroweak physics can assure this as
temperatures fall through the $T \sim 100$  GeV range if the Higgs field
gains a large condensate at a first order phase transition.  To avoid
the relaxation of baryon number back to zero in the broken phase, the
Higgs condensate $h$ must satisfy $h / T \gsim 1.1$
\cite{broken_nonpert}.  Such a strong phase transition is not
guaranteed, but it depends on the exact values of masses and
couplings.  In the standard model it does not occur; with the current
bound on the Higgs mass, $m_H > 95.5$  GeV \cite{Ellis_talk}, there is no
phase transition at all \cite{no_transition}.  
However, in the minimal supersymmetric
standard model (MSSM), {\em if} the mostly right-handed scalar top
quark (henceforth stop) is sufficiently light, then the phase
transition can be strong enough \cite{lots_of_people}.  (A light
left-handed stop is disfavored by its contribution to the precision
electroweak rho parameter.)  For this to occur, the right stop mass
parameter $m_U^2$ must be negative.  If mixing between right and left
stops is negligible, the mass of the light squark satisfies
$m_{\tilde{t}}^2 = m_t^2 + m_U^2$ at tree level, so the lightest squark
is lighter than the top quark.  If the left-handed stop is sufficiently
heavy, $m_Q \gsim 1$ TeV, then its radiative correction to the Higgs
boson mass is large enough to satisfy the experimental limit on $m_h$
even though the other top squark contributes negligibly to $m_h$.  This
appears to be the scenario for electroweak baryogenesis requiring the
least additional physics.

If $m_U^2$ is sufficiently negative (at tree level, if $m_U^2 < - (
g_s^2 / 6 y_t^2) m_H^2$), then there is a second, metastable minimum of
the electroweak potential, in which the stop field but not the Higgs
field condenses, and charge and color, but not SU(2)$_{\rm weak}$, are
broken.  At very high temperatures the only minimum of the free energy
is the symmetry restored one, but if $m_U^2$ is negative enough, this
charge and color breaking (CCB) minimum might become metastable at a
higher temperature than the conventional electroweak (EW) minimum.
This opens a qualitatively new scenario, first discussed by Kusenko,
Langacker and Segre \cite{KLS}, and more recently by B\"{o}deker,
John, Laine, and Schmidt \cite{John}, and 
Quiros {\it et al.}\ \cite{Quiros}.  The universe begins at high
temperatures in the symmetric phase.  As it cools, at some temperature
$T_{c1}$ the color breaking minimum appears, and shortly thereafter, at
$T_{\rm nuc 1}$, the universe converts to this phase via a
bubble-nucleation-driven first order phase transition.  Later, at a
temperature $T_{c2}$, the electroweak minimum becomes energetically
competitive with the symmetric phase, and at $T_{c3}$ its free energy
equals that of the color breaking minimum.  Finally, at some lower
temperature\footnote{We denote by $T_{\rm nuc 2}$ the temperature of
nucleation of electroweak bubbles from the symmetric phase, in the case
that color breaking does not occur first.}
 $T_{\rm nuc 3}$, the free energy difference between the minima becomes
sufficient to allow nucleation of bubbles of the EW phase out of the
CCB phase, and electroweak symmetry is broken and color symmetry
restored.\footnote{To be precise, a local, gauge symmetry is never
``broken'' in the sense of not being a symmetry of the ground state,
and no gauge invariant operator unambiguously distinguishes the
phases.  In fact, for some values of the couplings, the electroweak
``symmetric'' and ``broken'' phases are not distinct at all, and there
is no phase transition as the temperature is lowered
\protect{\cite{no_transition}}.  However, for the case at hand the
symmetry restored and broken phases have a good operational definition,
in terms of gauge invariant order parameters like $H^\dagger H$, and
there is no problem in distinguishing them.} Baryogenesis could occur
in this transition, which can be very strong.  It also has
a novel and  rich phenomenology; SU(3)$_{\rm color}$ is broken to SU(2)
in the color-breaking phase, and numerous mass eigenstates differ
between the phases.  The implications for baryogenesis have not been
studied in detail, although they could be very interesting.

But before studying them, we should first ascertain whether this
sequence of phase transitions can actually occur.  With the current,
very weak bounds on the physical stop mass, there is no problem making
$m_U^2$ negative enough; and there is a range of $m_U^2$ values where
color breaking would occur at a higher temperature, but the global
vacuum minimum would be the EW one.  But this does not guarantee that
the phase transition would have occurred cosmologically.  For the case
of the conventional electroweak phase transition, or the transition to
the color breaking phase mentioned above, there is always a temperature
where bubble nucleation becomes efficient, simply because the symmetric
phase eventually becomes spinodally unstable: the field can roll down
instead of tunneling.  On the other hand, for the CCB to EW phase
transition, both minima remain metastable down to $T=0$.  It may be
that, at some temperature, tunneling out of the CCB phase occurs
relatively quickly.  But it is also possible that the CCB phase may
satisfy Yoda's principle; ``Once you start down that dark path, forever
will it dominate your destiny.''  This paper attempts to determine
whether the nucleation rate is ever fast enough for escape from the CCB
minimum.

The efficiency of nucleation of the stable phase is controlled by the
action of the lowest saddle point configuration interpolating between
the two minima, in the Euclidean path integral with periodic time of
period $1 / T$ \cite{Afleck}.  At low temperature the time direction
can be approximated to be infinite, which allows one to recover the
results of Coleman and Callan \cite{Coleman}; in this limit the
critical action has the form $S = C / g^2$ and the tunneling rate is
therefore $\sim \exp(-C / g^2)$, where $g^2$ characterizes the coupling
constants of the theory and $C$ is a real number which depends on the
shape of the effective potential.  At high temperature, the saddle
point solution does not vary in the (very short) Euclidean time
direction at all, so the action is $S = E / T \sim m / g^2 T$, with $m
\sim g h$ a characteristic mass scale in the problem and $h$ the
separation of the minima in field space.  This leads to a nucleation
rate with the parametric form $\exp(-C'h/gT)$, where $C'$ is
another function of the shape of the potential.

\begin{figure}[t]
\centerline{\epsfxsize=6in\epsfbox{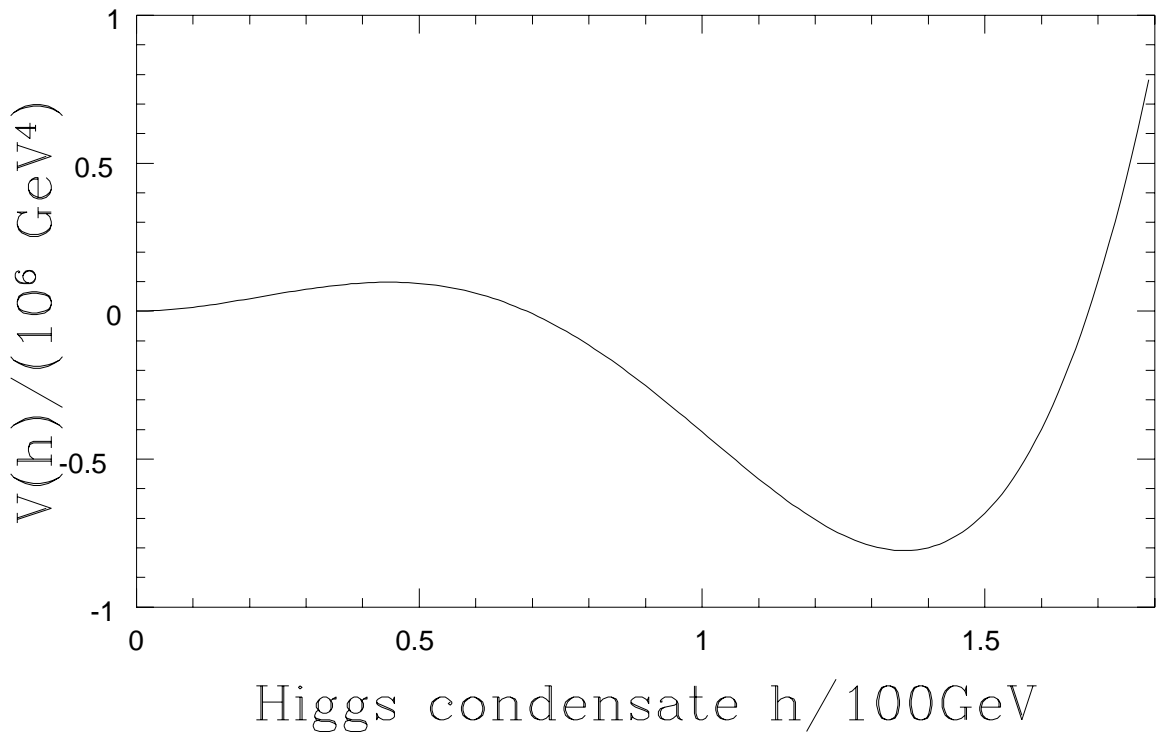}}
\caption{\label{fig1} The effective potential at the nucleation
temperature for the electroweak phase transition, in the standard
scenario where color breaking never occurs.
The barrier between phases is small compared to the difference in depths
of the phases.}
\end{figure}

If the two minima are nearly degenerate, then $C'$ is numerically large
and $C$ is even larger.  If one minimum is almost spinodally unstable,
$C$ and $C'$ can be small.  However the potential must come fairly
close to spinodal before $C'$ becomes as small as $O(1)$, which means
that in practice nucleation is very slow except near a spinodal
point.\footnote{One could imagine models with extra physics, for
instance cosmic strings coupling either to the Higgs or stop fields, in
which the phase transition could be stimulated by ``nucleation sites;''
here we will consider only the case with no additional exotic physics.}
Figure \ref{fig1} illustrates this point by showing the shape of the
potential for the Higgs field at the temperature $T_{\rm nuc 2}$ where
the Higgs phase nucleates out of the symmetric phase, at a value of
parameters for which color breaking does not occur.  One notices that
the height $\epsilon$ of the ``bump'' separating the two phases is
small compared to the free energy difference $\Delta V$ between them.
This is typical, and the value of of $\epsilon/\Delta V$ required to
make the phase transition complete is even smaller if the strength of
the transition (measured by $\langle h \rangle/T$) is increased.

The tunneling rate from the CCB to the EW minimum behaves similarly,
but unlike the pure electroweak transition, its bump height $\epsilon$
need not go to zero.
Moreover the phase transition is strong; $\langle h \rangle/T$ becomes
quite large by the time the critical temperature for this second
transition is reached.  This requires a very small $\epsilon/\Delta V$,
and we are right to wonder whether that will be achieved.  Hence,
our aim must be to determine not {\it when} the CCB phase tunnels to
the EW phase, but whether it can {\em ever} do so, on cosmologically
relevant time scales.  In Section \ref{general_sec} we make some rough
estimates to determine what region of parameter space has the fastest
tunneling rate.  The construction of the finite temperature effective
potential is discussed in Section \ref{Veff_sec}.  The details of how
we compute the tunneling rate follow in Section \ref{detail_sec}, while
the technical details of the calculation of the critical bubble shape
and action are postponed to Appendix \ref{AppendixA}, and the
renormalization group analysis needed to find the couplings of the tree
level potential is described in Appendix \ref{AppendixB}.  We conclude
that the nucleation is too slow for EW bubbles ever to percolate, for
any physically allowed values of the MSSM couplings.

\section{Rough estimates and the choice of parameters}
\label{general_sec}

Before constructing the full effective potential, it is useful to
discuss a rough approximation which can give much analytic insight into
the dependence of the tunneling rate on the many unknown parameters of
the MSSM.  For this purpose, the most important terms in the
approximate potential are those which determine the critical
temperatures $T_{c1}- T_{c3}$, as well as the height of the barrier
between the color-broken and electroweak phases.  These are precisely
the quadratic and quartic couplings that appear in the zero-temperature
effective potential, but with coefficients that now depend on temperature.
A more accurate approximation would require the temperature-induced
cubic terms as well, but these are not necessary for the analysis of 
this section.  Only in the following section will we present the full
effective potential with all terms included.

\subsection{Preferred values of the couplings}

At tree level and in the absence of squark mixing, 
and assuming the $A^0$ boson mass is large so that only one linear
combination of the two Higgs doublets is light, 
the effective potential for the MSSM is
\beq
\label{tree_potential}
V(h,s) = - \frac{\mu^2_h}{2} h^2 - \frac{\mu^2_s}{2} s^2 
	+ \frac{\lambda_h}{4} h^4 + \frac{\lambda_s}{4} s^4
	+ \frac{\lambda_y}{4} h^2 s^2 \, .
\eeq
Here $h$ denotes the 
Higgs condensate and $s$ the right stop condensate, both normalized
as real fields. 
The coupling between the $h$ and $s$ fields is written as $\lambda_y$
because of its relation to the top quark Yukawa coupling $y$:
$\lambda_y = 
y^2 \sin^2\beta$, where $\beta$ is defined by the
ratio of the two Higgs field VEV's, $\tan\beta = \langle H_2\rangle
/\langle H_1 \rangle$.
At leading order in couplings, and in the high temperature expansion, 
the thermal corrections to this potential take the form of an 
irrelevant additive
constant, plus thermal corrections to the mass
parameters,
\beq
\mu^2_h(T) = \mu^2_h - c_h T^2 \, , 
\qquad \mu^2_s(T) = \mu^2_s - c_s T^2 \, .
\eeq
The values of $c_h$ and $c_s$ depend on which degrees of freedom are
light, as well as their couplings.

Presently we will relate the masses and couplings of our approximate
potential to the parameters of the MSSM.  First, however, we would like
to show how the tunneling rate depends on the $\mu^2_i$ and $\lambda_i$.
The goal is to identify those values which give the maximum tunneling
rate, which in turn will help us choose the parameters of the MSSM which
are most favorable to tunneling out of the CCB phase.

First we consider the locations and depths of the two minima.
The Higgs and stop minima, $h_0$ and $s_0$, are characterized by
\beqa
h_0^2 & = & \frac{\mu^2_h}{\lambda_h}\, ,\qquad
s_0^2  =  \frac{\mu^2_s}{\lambda_s} \, , \\
V(h_0 , 0)  & = & - \frac{\mu^4_h}{4 \lambda_h} \, ,\qquad
V(0 , s_0)   =  - \frac{\mu^4_s}{4 \lambda_s}\, .
\eeqa
Therefore a minimum is deeper if the relevant $\mu^2$ is larger and the
relevant $\lambda$ is smaller.  Since the best case for tunneling is
when the CCB minimum is as shallow as possible compared to the EW one,
tunneling prefers small $\lambda_h$ and $\mu^2_s$ but large $\lambda_s$
and $\mu^2_h$.

Next we examine the critical temperatures.
At tree level, the temperatures $T_{c1}$, $T_{c2}$ where the 
symmetric phase 
destabilizes in favor of the CCB or EW phase, respectively, are
\beq
T_{c1}^2 = \frac{\mu^2_s}{c_s} \, , \quad T_{c2}^2 = 
	{\mu^2_h\over c_h} \, .
\eeq
We require $T_{c1} \geq T_{c2}$ to get the right sequence of symmetry
breakings.  A large value for $T_{c1}$ conflicts with the need to
minimize $\mu^2_s$, so the optimal choice is for the phase transition
temperatures to be almost the same, $T_{c1} \simeq T_{c2}$.  The ratio
$\mu^2_s / \mu^2_h$ equals $c_s / c_h$ in this case; so tunneling is favored
by a small thermal correction to the stop mass, $c_s$, but a large thermal
correction to the Higgs mass, $c_h$.

We should also consider the size of the 
barrier between the minima.  It is highest
for large values of $\lambda_y$, because the large positive $s^2 h^2$
term in the potential prevents the two fields from simultaneously having
large expectation values.  To see this, let us find the saddle point of
the potential between the two minima.  Fixing $s^2 / h^2 = R$, then
minimizing $V$ with respect to $s^2$ at fixed $R$, gives
\beqa
R &\equiv& \frac{s^2}{h^2} = {2 \mu^2_s \lambda_h - \mu^2_h \lambda_y \over
2 \lambda_s \mu^2_h - \lambda_y \mu^2_s} \nonumber\\
 \quad \Rightarrow \quad
V(R) &=& - \frac{1}{4} \frac{ (\mu^2_h + R \mu^2_s)^2}{\lambda_h + R \lambda_y
	+ R^2 \lambda_s} \, .
\eeqa
The saddle point is the maximum of $V(R)$ over positive values of $R$.
Such a saddle exists if the inequalities
\beq
\label{stab_cond}
\frac{\lambda_y}{2\lambda_h} > \frac{\mu^2_s}{\mu^2_h} >
\frac{2\lambda_s}{\lambda_y}
\eeq
hold; if not then either the CCB or the EW ``minimum'' is not a local
minimum but a saddle point.  This does not happen for any physically
allowed parameters which give $T_{c1} > T_{c2}$, 
so in practice there is always a saddle.  Its depth is
\beq
\label{saddle_value}
V({\rm saddle\ point}) = - \frac{\mu^2_s \mu^2_h \lambda_y - \mu^4_h \lambda_s
	- \mu^4_s \lambda_h}{\lambda_y^2 - 4 \lambda_s \lambda_h} \, .
\eeq
The inequalities (\ref{stab_cond}) imply that both numerator and
denominator of Eq.\ (\ref{saddle_value}) are positive, so that
its overall value is negative.  If we hold $\lambda_h$, $\lambda_s$,
$\mu^2_s$, and $\mu^2_h$ fixed, 
the saddle energy is lower for smaller values of $\lambda_y$,
rising to zero as $\lambda_y \rightarrow \infty$.

Thus we can summarize our study of the simplified effective potential by
the observation that tunneling is easiest to achieve for small
$\lambda_h$, large $\lambda_s$, small $\lambda_y$, and large $c_h / c_s$.

\subsection{Relation to MSSM parameters}
\label{rmpsubsect}

Next we will examine the physical bounds on these variables and consider
the choices for SUSY breaking masses and other MSSM parameters which
optimize tunneling from the CCB phase. 

\begin{figure}[t]
\centerline{\epsfxsize=4in\epsfbox{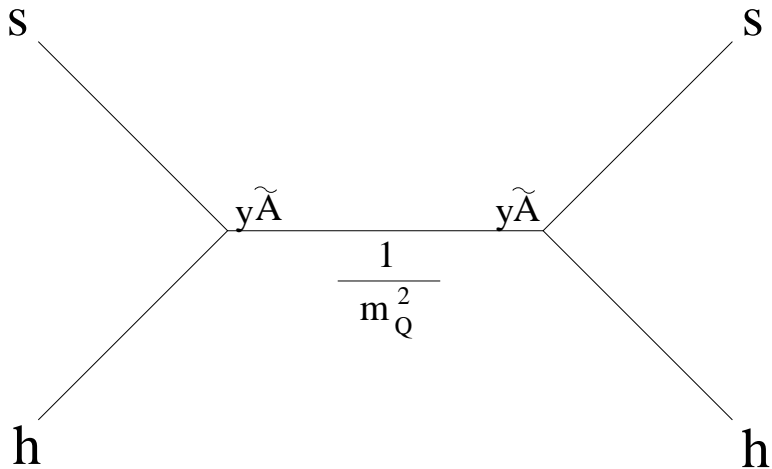}}
\caption{\label{shift_fig} A tree level correction caused by
a heavy left stop at nonzero mixing.  It effectively lowers the quartic
coupling between the Higgs and stop fields.}
\end{figure}

We begin with $\lambda_y$.  By introducing mixing between the left-
and right-handed stops, it is possible to tune $\lambda_y$ to any
desired value smaller than its zero-mixing value, which at tree level is  
$y^2 = 2 m_t^2 / h_0^2$.  This is true no matter how heavy the heavy 
(left) stop
is.  To see this, consider the tree level potential for the $h$
and $s$ fields, but also allowing for a left stop condensate $Q$.  
The terms in the potential which depend on the $Q$ field are 
\beq
V(h,s,l) = - \frac{\mu^2_h}{2} h^2 - \frac{\mu^2_s}{2} s^2 
	+ \frac{m_Q^2}{2} Q^2 + \frac{y \sin \beta \tilde{A}}
	{\sqrt{2}} shQ +
	( Q^4 {\rm \; and \;} Q^2 h^2,Q^2 s^2 {\rm \; terms}) \, .
\eeq
Here $\sin \beta \tilde{A} = \mu \cos \beta 
+ A_t \sin \beta$ is the trilinear
coupling between the right stop, left stop, and Higgs fields, which is a
free parameter in the MSSM.  The
potential is minimized with respect to $Q$ at fixed $s$ and $h$ by 
$Q = (-y \sin \beta \tilde{A}/m_Q^2 \sqrt{2}) sh$, 
up to corrections suppressed by
powers of $h^2/m_Q^2$ or $s^2/m_Q^2$.  At this field value the $Q$
dependent contributions sum to $(- y^2 \sin^2 \beta 
\tilde{A}^2/4 m_Q^2) s^2 h^2$.
This is equivalent to a shift in the value of $\lambda_y$,
\beq
\label{tree_lamy}
\lambda_y({\rm effective}) = 
	y^2\sin^2\beta\left( 1 - \frac{\tilde{A}^2}{m_Q^2} \right)
	+ \sfrac13 g'^2\cos 2\beta \, .
\eeq
This shift can also be understood as a consequence of the diagram shown in
Figure \ref{shift_fig}.  If we allow $\tilde{A}^2/m_Q^2$ to be of order
unity the effect is significant, while the corrections of order
$h^2/m_Q^2$ or $s^2/m_Q^2$ which we neglected only give high dimension
operators suppressed by powers of $m_Q^2$.  We ignore them in what
follows.

The reduction of $\lambda_y$ is the only tree level effect of squark
mixing, apart from the small nonrenormalizable operators.  By varying
$\tilde{A}^2/m_Q^2$ we can therefore tune $\lambda_y$ to be any value
lower than its zero-mixing value.  However, there is an experimental
constraint; a top squark lighter than $85$  GeV is ruled out
\cite{Ellis_talk}.  This puts an upper bound on the permissible value
of $\tilde{A}^2/m_Q^2$.

Although we concluded the previous subsection by saying that making
$\lambda_y$ small should be advantageous for tunneling, doing so also
has a cost;  by diminishing the coupling between the Higgs and stop
fields, it also reduces $c_h$, more so than $c_s$.  This is because a
triplet ($N_c$) of thermal squarks contribute to $c_h$ via the
$\lambda_y$ interaction, whereas only a doublet of thermal Higgs bosons
contribute to $c_s$ by the same interaction.  Moreover $c_s$ is already
larger than $c_h$, so the fractional change to $c_h$ is even worse.
This shift in the thermal masses goes in the wrong direction so far as
the CCB to electroweak tunneling is concerned.  Additionally, a nonzero
value of $\tilde{A}$ changes other radiative corrections.  Because
of these complications, we do not try to predict the optimum value of
$\tilde{A}$; rather we will treat $\tilde{A}^2/m_Q^2$ as a free
parameter and search for the most favorable value, within the range
permitted by the experimental bound on the physical squark mass.

Next consider $\lambda_s$, $\lambda_h$, $c_s$, and $c_h$.  In the
supersymmetric limit the quartic couplings are given in terms of
the gauge couplings ($g'$, $g$, $g_s$) and $\beta$:
\beq
\label{tree_relations}
\lambda_h = \left( \frac{g^2 + g'^2}{8} \right) \cos^2 2 \beta \, ,
	\qquad \lambda_s = \frac{g_s^2}{6} + \frac{2 g'^2}{9} \, ,
\eeq
but both relations, as well as Eq.\ (\ref{tree_lamy}), 
are violated below the mass thresholds of heavy
particles.  The most important corrections are those which 
involve $g_s$ and $y$.  We 
will systematically include all such corrections to $\lambda_h$,
$\lambda_y$, and
$\lambda_s$.  However we will be less careful with the much smaller
corrections of order $g^4$ and will
drop the bottom and tau Yukawa couplings altogether.

Among the particles assumed to be heavy, whose loop effects change the
tree level relations (\ref{tree_relations}), the squarks of the first two
generations and the right sbottom are important because of their strong
interactions.  Above their mass threshold they make the running coupling
$g_s^2(\overline{\mu})$ larger in the ultraviolet, but they also make
$\lambda_s(\overline{\mu})$ run
faster in the same direction; thus their absence, when the renormalization
scale falls below their mass threshold, causes the infrared value of
$\lambda_s$ to be larger than its supersymmetric value; at one loop the
difference is
\beq
\delta\left(\frac{6 \lambda_s}{g_s^2} - 1
\right)_{\rm squarks} = 
	\frac{2}{3} \times \frac{g_s^2}{16 \pi^2} \sum 
	\left( \ln \frac{m_{\tilde{q}}}{\overline{\mu}} + O(1) \right) \, .
\eeq
The term denoted by ``$O(1)$'' is actually zero in the $\overline{\rm DR}$
renormalization scheme, which we use, so we shall henceforth drop it.  The
sum is over flavors and chiralities, 9 in all.  The heavier these squarks
are, the easier is the nucleation; hence we take them to have masses of 10
TeV, since larger values may not be consistent with low-energy
SUSY from the standpoint of naturalness.  Since $g_s^2$ runs significantly
between $10$ TeV and the electroweak scale, a renormalization group
analysis is indispensable for determining the correct relation between
$\lambda_s$ and $g_s^2$.  In fact we will perform a renormalization
group analysis for all the scalar couplings, but in this section we
just present the one loop results to see which way couplings are
modified, so we can choose the optimal parameter values.

\begin{figure}[t]
\centerline{\epsfxsize=5in\epsfbox{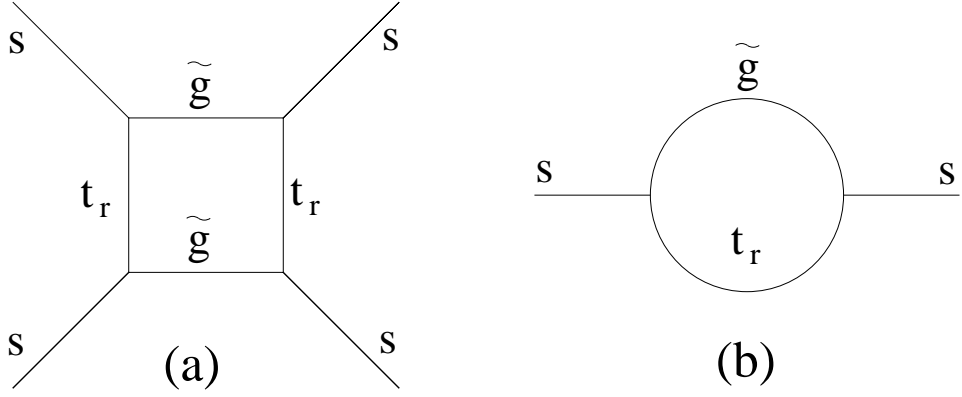}}
\caption{\label{gluino_loops} Gluino loop contributions to
(a) the quartic coupling $\lambda_s$ and (b) the light squark
thermal mass and wave function renormalization.}
\end{figure}

Continuing the analysis of $\lambda_s/g_s^2$, we
next consider the effects of gluino loops, such as the diagrams in
Figure \ref{gluino_loops}.  These correct $\lambda_s$, and also
contribute to the light squark thermal mass coefficient $c_s$ if the
gluino is not heavy compared to the weak scale.  The latter
contribution is a function of $m_{\tilde{g}} / T$:
\beq
\label{therm_mass}
\delta (c_s)_{\rm gluino} 
	= \frac{g_s^2}{9} \left[ \frac{\pi^2}{12} 
	\int_{m_{\tilde{g}}/T}^\infty 
	\frac{\sqrt{x^2 - (m_{\tilde{g}}/T)^2}}{e^x + 1}\, dx \right] \, .
\eeq
The term in brackets goes to 1 at small $m/T$ and behaves like $e^{-m/T}$ 
for large $m/T$.  In the former case, the correction to $c_s$
is quite large and tends to inhibit tunneling from the CCB vacuum.  Thus
we should try to suppress this thermal mass by taking the gluino to 
be heavy.  However, the gluino also shifts $6 \lambda_s/ g_s^2$, 
\beq
\delta\left(\frac{6 \lambda_s}{g_s^2} - 1\right)_{\rm gluino} = 
	-\frac{68}{3} \times \frac{g_s^2}{16 \pi^2} 
	\ln \frac{m_{\tilde{g}}}{\overline{\mu}}
	 \, .
\eeq
The shift is large and unfavorable for tunneling; it is minimized by
making the gluino light.  The best value
for $m_{\tilde{g}}$ is around $600$  GeV, which is as small as it can
be while still avoiding a 
substantial correction to the thermal stop mass.  Later we will
show numerically that this value really is optimal.

\begin{figure}[b]
\centerline{\epsfxsize=5in\epsfbox{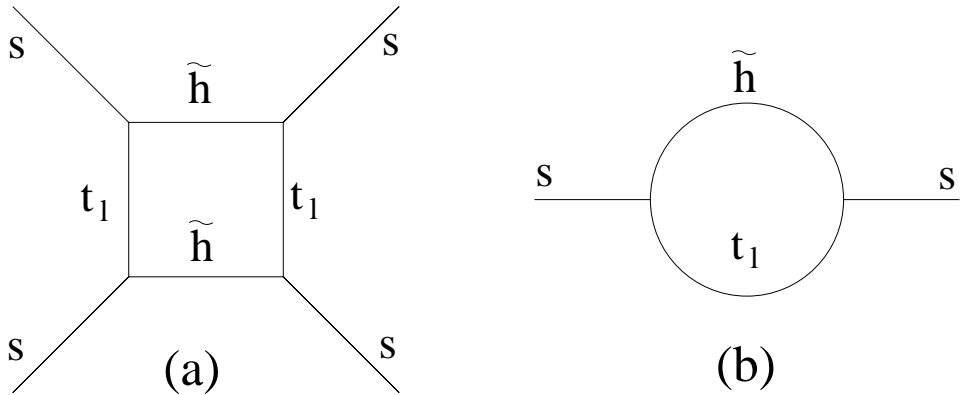}}
\caption{\label{higgsino_loops} Higgsino loop contributions to
(a) the quartic coupling $\lambda_s$ and (b) 
the light squark thermal mass and wave function renormalization.}
\end{figure}

Similarly, Higgsino ($\tilde h$) loops shift the stop quartic coupling
and thermal mass through the diagrams of Figure \ref{higgsino_loops}.
The correction to the thermal mass, for Higgsinos that are
light enough to be present in the thermal background, is
\beq
\delta(c_s)_{\rm Higgsino}  =  \frac{y^2}{12} \, .
\eeq
Since we want to minimize $c_s$, this gives some preference for a 
heavy Higgsino.  However, the 
shift in $\lambda_s/g_s^2$ has the form
\beq
\delta\left(\frac{6 \lambda_s}{g_s^2} - 1\right)_{\rm Higgsino} 
	= \frac{-24 y^4 + 8 y^2 g_s^2}{16 \pi^2 g_s^2}
	\ln \frac{m_{\tilde{h}}}{\overline{\mu}}\, .
\eeq
Since the coefficient is negative, the need to maximize $\lambda_s$
makes this favor lighter Higgsinos.  We infer that, like gluinos,
the Higgsino should also be of intermediate weight for fastest tunneling.

It remains to determine $\tan \beta$, the mass of the heavy Higgs bosons
$m_{A^0}$, and the mass $m_Q$ of the third generation left squark doublet,
including the left stop.  The contribution of the heavy Higgs bosons
to $c_h$ turns out to be negative, and there is a positive
contribution to $c_s$ due to their Yukawa coupling, which is however
suppressed by $\cos^2 \beta$.  For these reasons it is best to make them
heavy.  They also give radiative corrections which make $\lambda_s$
larger as $m_{A^0}$ becomes heavier.  The form is complicated because
there is another trilinear coupling between the heavy Higgs, the right
stop, and the left stop.  To avoid this complication and because a heavy
$m_{A^0}$ is preferred anyway, we take the $A^0$ mass to be degenerate
with the left stop squark mass.

Now consider $m_Q$ and $\tan \beta$.  We want $\tan \beta$ to be small
to minimize $\lambda_h$, and for the same reason it would be
advantageous to make $m_Q$ light.  However this desire is constrained
by the need to make the physical Higgs boson heavier than the limit
from direct experimental searches: $m_h > 95.5$  GeV for a
standard-model-like Higgs boson, to which the MSSM Higgs boson reverts
in the limit of large $m_{A^0}$ \cite{Ellis_talk}.  $m_h$ can be made
sufficiently heavy either by making $\tan \beta$ or $m_Q$ large.
The question is
therefore which parameter does less harm to the phase transition if it
is increased.  To answer this, we must consider the radiative
corrections from the heavy squark to each coupling (assuming $m_{A^0} =
m_Q$):
\beqa
\!\!\!\!\!\!\!
\delta\left(\frac{6 \lambda_s}{g_s^2} - 1
	\right)_ {\rm left \ stop} & = & 
	\frac{1}{16 \pi^2 g_s^2}
	\left( 12 y^4\left[ 1 + 2 \frac{\tilde{A}^2}{m_Q^2} 
	- \frac{\tilde{A}^4}{m_Q^4} \right] - 8 y^2 g_s^2 
	+ (4/3) g_s^4 \right) \ln \frac{m_Q}{\overline{\mu}}\\
\delta (\lambda_h)_{\rm left \ stop}
	& = & \frac{1}{16 \pi^2} \left( 
	3 y^4 \sin^4 \beta \left[ 1 +2 \frac{\tilde{A}^2}{m_Q^2}
	- \frac{\tilde{A}^4}{m_Q^4}  
	\right] - 12 y^2 \lambda_h \sin^2 \beta 
	\right) \ln \frac{m_Q}{m_t} \, .
\eeqa
The contribution to $\lambda_s$ is positive and therefore favorable to
nucleation.  The best combination is therefore to make $\tan \beta$
small and $m_Q$ just large enough to satisfy the Higgs mass limit; this
maximizes $\lambda_s$ over parameter values where $\lambda_h$ is at its
experimental lower limit.  As the expressions show, the contribution to
$\lambda_h$ is larger if there is mixing.  We either take $\tan
\beta=2.5$ and fix $m_Q$ to be the minimum value needed to satisfy the
limit on 
the Higgs mass, or if the resulting value of $m_Q$ exceeds $10$~TeV, we
take $m_Q=10$~TeV and $\tan \beta$ the smallest value which satisfies
the Higgs mass bound.  Using the one loop expressions above, the value
of $m_Q$ need never be $10$~TeV, but in a renormalization group
analysis, because $y(\overline{\mu})$ is less than the tree value for
large $\overline{\mu}$, the squark mass must be heavier.

Finally we must choose masses for the Wino, the Bino, and the
sleptons.  For simplicity we omit the sleptons altogether, since their
contributions are small.  We cannot do the same for the Wino and the
Bino because the lightest supersymmetric partner must be neutral;
something needs to be lighter than the right stop.   
Since the Higgsino has already been designated as
moderately heavy, some linear combination of the neutral Wino and
Bino must be the lightest supersymmetric particle.  We have chosen to
make the Winos light; but we have also checked that our 
results are quite insensitive to the values of the neutralino masses.

This completes our discussion of the choice of parameters.  We have
analytically predicted the most favorable range for every parameter
except the mixing parameter $\tilde{A}^2 / m_Q^2$, which must therefore
be varied to find the optimal value for tunneling.  Of course, we will
also verify the predictions of this section by varying each parameter
from its optimal value.

It is not clear whether any of our choices can be motivated by a
specific model of supersymmetry breaking.  But this is not the point;
we want to identify the optimal values of all MSSM parameters to obtain
the largest possible tunneling rate.   Since the rate turns out to be
too small, any further restrictions on the
space of SUSY parameters from model building considerations will only
strengthen our conclusions.

\section{The Effective Potential}
\label{Veff_sec}

Here we discuss the effective potential we use, paying particular
attention to the choice for scalar couplings and to the rather
complicated mass matrices which occur when there are two condensates.
The first step is to find the mass eigenvalues of all particles which
run in loops, as a function of the arbitrary background fields whose
effective potential is sought.  In the present case, we must find the
masses as functions of $h$ and $s$, the Higgs and squark fields.  This
task is complicated by the large degree of mixing between many
different flavor eigenstates when both fields are nonzero, but since we
will numerically diagonalize all mass matrices, this is not a problem
in practice.  

Once the mass eigenvalues are known, the one-loop
potential can be expressed as
\beq
   V_{\rm eff}(h,s) = V_{\rm tree} + V_{\rm c.t.} +
	V_{1,\rm vac} + V_{1,\rm therm} \, .
\eeq
Here $V_{\rm tree}$ is the tree-level potential, Eq.\
(\ref{tree_potential}), with 
couplings and masses to be specified presently in great detail,
$V_{\rm c.t.}$ is a counterterm potential which could be considered part
of $V_{\rm vac}$ but is kept separate for convenience,
and the remaining terms are the one-loop vacuum and thermal
contributions.  
The vacuum part is the Coleman-Weinberg potential
at a renormalization scale $\overline{\mu}$,
\beq
	V_{1,\rm vac}(h,s) = {1\over 64\pi^2}\sum_i \pm m^4_i(h,s)
	\left(\ln{m^2_i(h,s)\over\overline{\mu}^2} - {3\over 2}\right),
\label{CW}
\eeq
with $\pm$ being $+$ for bosons and $-$ for fermions in the sum over
species.  Each real scalar or physical gauge boson polarization, and
each helicity of a Weyl fermion counts as one state.  The constant
$3/2$ would be $5/6$ for gauge boson contributions in the
$\overline{\rm MS}$ scheme, but in $\overline{\rm DR}$, which we adopt,
all particles have $3/2$.  
The thermal part of the potential, before
resummation of thermal masses, is given by
\beq
\label{V_therm}
	V_{1,\rm therm}(h,s) = {T^4\over (2\pi)^3}\sum_i \pm 
	\int d^{\,3}p\, \ln\left(1\mp e^{-\sqrt{p^2 + m^2_i(h,s)}/T} \right).
\eeq
This is sometimes approximated by its high-temperature expansion, but 
we also need the correct values at low temperatures.  A convenient
analytic form which is accurate at both high and low $T$ is given in 
ref.\ \cite{CL}.
To improve convergence of the perturbation expansion at finite
temperature, it is important to resum the thermal masses of the particles
by replacing $m^2_i$ with $m^2_i + c_i T^2$ in Eq.\ (\ref{V_therm}).
The form $c_i T^2$ is only valid in the high temperature limit, so we
will instead use a more exact determination, to be described below,
for the thermal masses of the Higgs and squark fields.

\subsection{Definition of $V_{\rm tree}$ and $V_{\rm ct}$}

To fully define $V_{\rm eff}$, we must specify the masses and couplings
in $V_{\rm tree}$, and which particles appear in the sum over species
of the one-loop part. The two questions are related, since the loop
effects of any particles not explicitly appearing in the sums should be
directly incorporated into the couplings of $V_{\rm tree}$.  We have
chosen to exclude the following particles from the sum over species:
first and second generation squarks, the left-handed stop and sbottom,
and the heavy Higgs bosons.  Sleptons are entirely omitted, and light
quarks and leptons are counted only insofar as they affect the thermal
(Debye) mass coefficients $c_i$.  All other particles appear in the
summations:  the gauge bosons, gauginos, neutralinos, Higgsinos, top
quark, right-handed stop, and light Higgs boson.  In addition, the
color-component of the left-handed bottom quark in the color-breaking
direction mixes with the charged Higgsino $\tilde h_2^+$ in the
presence of the squark condensate, so it must also be included.  The
decision as to whether to include particles explicitly is based upon
how large a contribution they make to $V_{1,\rm
therm}$, which contains terms of the form $T m^3_i$ at high
temperatures.  Such a dependence on the fields cannot be reproduced by
the quadratic and quartic terms in $V_{\rm tree}$.  On the other hand,
particles with masses much greater than $T$ are negligible in $V_{1,\rm
therm}$, and their contributions to $V_{1,\rm vac}$ can be expressed 
as purely quadratic and quartic terms for field values much less than
the large masses.

Our choice for $V_{\rm tree}$ is as follows.  For the quartic scalar
couplings $\lambda_h$, $\lambda_s$, and $\lambda_h$, we use their
values at the $\overline{\rm DR}$ renormalization point
$\overline{\mu}$, in the effective theory in which all heavy squarks
and the gluino and Higgsino have been integrated out.  These are
determined by a renormalization group (RG) analysis, which can be found
in Appendix \ref{AppendixB}.  Applying an RG analysis is important to
get accurate values of the scalar couplings because $\alpha_s$ is not
{\em very} small and because we have taken some masses to be very
large, leading to large hierarchies and large logarithms.  The
difference between performing the RG analysis and simply enforcing the
SUSY relations between couplings at the scale $\overline{\mu}$ is of
order a $20\%$ shift in scalar self-couplings, and the difference
between doing an RG analysis and a simple one-loop match is smaller but
still not negligible.

The result of the analysis is that the coupling $\lambda_y$ is
substantially lower than its tree value, $\lambda_y(\overline{\mu})
\simeq 0.71$ rather than 1; this is partly because of the QCD correction
between the Yukawa coupling and the top quark mass and partly because of a
large downward correction from the gluino.  The value of $\lambda_s$ is
surprisingly close to its SUSY value using $g_s^2$ at the $Z$ pole;
typically $\lambda_s \simeq 0.24$.  This is because of an approximate
cancellation between positive contributions from the gluino and Higgsino,
which are naturally large, and negative contributions from the heavy
squarks which we have enhanced by choosing these squarks to be
extremely heavy.  The Higgs coupling $\lambda_h$ is expected to receive
large radiative corrections, but they are not as large as usually
expected, because of the threshold correction to the Yukawa coupling and
because the Yukawa coupling gets weaker in the UV.  As a result the left
stop must be very heavy and $\tan \beta$ must be about 3 to reach
the experimental limit on the Higgs mass, unless there is mixing.

Note that both the correction to $\lambda_y$ and the slower running of
$\lambda_h$ are bad for the ``usual'' scenario in which only the
electroweak phase transition occurs.  The lower $\lambda_y$ weakens the
electroweak phase transition, narrowing the permitted range of
parameters; and the smaller corrections to $\lambda_h$ require
a larger hierarchy between the left and right stop masses to satisfy the
experimental Higgs mass limit, which increases the amount of tuning
needed in setting the SUSY breaking parameters.

Having chosen the scalar self-couplings in the tree potential, we now
specify the mass parameters.  The value of $\mu^2_h$ is chosen so the
minimum of the tree potential occurs at $v=246$~ GeV, and $\mu^2_s$ is
an input variable.

Next we consider the counterterm potential, 
 $V_{\rm c.t.}$.  The tree and one loop effective
potentials just described double-count the influence of any heavy
particle left out of part of the RG evolution but included in
Eq.\ (\ref{CW}), which in our case means the gluinos and the
Higgsinos.  Hence we need to subtract off the extra
contribution to the quartic coupling.  Also, Eq.\ (\ref{CW})
generates potentially large finite corrections to the Higgs and squark
masses, and we must include counterterms to absorb these.  The full
counterterm contribution is then
\beqa
V_{\rm c.t.} & = & - \frac{1}{2} \delta \mu_h h^2
	- \frac{1}{2} \delta \mu_s s^2
	- \frac{\delta \lambda_s}{4} s^4
	- \frac{\delta \lambda_h}{4} s^2 h^2 \, , \\
\delta \lambda_s & = & - \frac{44}{9} g_s^4 \log 
	\frac{m_{\tilde{g}}}{\overline{\mu}}
	-4 y^4 \log \frac{m_{\tilde{h}}}{\overline{\mu}} \, , \\
\delta \lambda_y & = & - \frac{32}{3} g_s^2 y^2 \sin^2 \beta \log 
	\frac{m_{\tilde{g}}}{\overline{\mu}}
	-4 y^4 \sin^2 \beta 
	\log \frac{m_{\tilde{h}}}{\overline{\mu}} \, .
\label{lyeq}
\eeqa
The coefficients in Eq.\ (\ref{lyeq}) come from Eq.\ (\ref{CW})
and the expression for the fermion mass matrix, to follow in
Eqs. (\ref{horrid1}) and (\ref{horrid2}) below.

The Higgs mass counterterm is fixed by the condition that the tree-level
minimum of
the vacuum potential should not be shifted, 
\beq
-\delta\mu_h\, v + {\partial V_{1,\rm vac}\over \partial h}(v,0)
	=0 \, .
\eeq
For the squark mass term, we choose the corresponding mass
counterterm $\delta\mu_s$ to cancel the one-loop contribution to the
curvature at the symmetric point:
\beq
-\delta\mu_s + {\partial^2 V_{1,\rm vac}\over \partial s^2}(0,0)
        =0 \, ,
\eeq
so the parameter $\mu_s$ retains its interpretation as the negative
curvature of the potential at the origin.

\subsection{Field-dependent masses}

We are now ready to turn our attention to the one-loop contributions.
The main challenge here is to find the mass eigenstates in the regions
where $h\neq 0$ and $s\neq 0$, where the mass matrices can become rather
large due to mixing between states which remain separate in the more 
familiar situation where $s=0$.  The simplest example is the Higgs boson,
$h$, and the squark component in the color-breaking direction, $s$.
Their $2\times 2$ mass matrix is
\beq
	{\cal M}^2_{h,s} = \left( \begin{array}{cc}
	\lambda_h (3 h^2 - v^2) + \sfrac12\lambda_y s^2 &
	\lambda_y h s \\ \lambda_y h s & -\mu^2_s + 3\lambda_s s^2
	+ \sfrac12 \lambda_y h^2 \end{array} \right)\, .
\eeq

Next in complexity are the gauge bosons.  Because both $s$ and $h$ carry
hypercharge, 
there is mixing between the three kinds of gauge
bosons when both fields are nonzero.  Take the 
color-breaking direction to be $a=3$ in the fundamental representation
of SU(3) with color indices $a$.  Then the mixing takes place between
the $B$, $W_3$, and $A_8$ gauge bosons (each having three polarization
states), with mass matrix
\beq
\label{gb_matrix}
	{\cal M}^2_{g.b.} = \left( \begin{array}{ccc}
	\sfrac14\, g'^2 h^2 + \sfrac49\, g'^2 s^2 & -\sfrac14\, g g'h^2 &
	 -\sfrac{2}{3\sqrt{3}}\, g' g_s s^2 \\
	-\sfrac14\, g g' h^2 & \sfrac14\, g^2 h^2 & 0 \\
	-\sfrac{2}{3\sqrt{3}}\, g' g_s s^2 & 0 & \sfrac13\, g_s^2 s^2
	 \end{array} \right)\, .	
\eeq
In fact only two of the eigenvalues of (\ref{gb_matrix}) are nonzero,
since there is still one linear combination of generators which gives
an unbroken $U(1)$ symmetry, even when both VEV's are present.  There
is also an unbroken SU(2) generated by $A_1$, $A_2$, $A_3$, so these
gluons remain massless.  The four gluons $A_4$-$A_7$ remain unmixed, but
get a mass
\beq
	m_g = \sfrac14\, g_s^2 s^2.
\eeq

The most baroque sector is that of the fermions.  When $s\neq 0$, there
is mixing between the charginos and the component of the left-handed
bottom quark in the color breaking direction, $b_L^3$.  There is also
mixing between top quarks, five of the gluinos, and all the
neutralinos.  These can be described by $5\times 5$ and $15\times 15$
Majorana mass matrices.  The chargino-$b_L$ mass matrix, in the basis
$\widetilde W^-$, $\widetilde W^+$, $\tilde h^-_1$, $\tilde h^+_2$, 
$b^3_L$, is
\beq
	{\cal M}_{\chi^\pm, b_L} = \left( \begin{array}{ccccc}
	0 & m_2 & 0 & \sqrtwo\eta_2 & \\
	m_2 & 0 & \sqrtwo\eta_1 & 0 & \\
	0 & \sqrtwo\eta_1 & 0 & \mu & \\
	\sqrtwo\eta_2 & 0 & \mu & 0 & -\sfrac{y}{\sqrt{2}}\, s \\
	 & & & -\sfrac{y}{\sqrt{2}}\,s & 0 
	 \end{array} \right)\, ,
\label{horrid1}
\eeq
where we define
\beqa
	\eta_1 &=& \sfrac{1}{2}\, g h \cos\beta; 
	\quad \eta_2 = \sfrac{1}{2}\,g h \sin\beta;\nonumber\\
	\eta'_1 &=& \sfrac{1}{2}\, g' h \cos\beta; 
	\quad \eta'_2 = \sfrac{1}{2}\, g' h \sin\beta\, .
\eeqa
The spectrum is that of two Dirac fermions and one massless one.  For
the top-gluino-neutralino mass matrix we have, in the basis $t_L$,
$t_R^c$, $\tilde g$, $\widetilde B$, $\widetilde W^0$, $\tilde h_1^0$,
$\tilde h_2^0$,
\beq
	{\cal M}_{t,\tilde g,\chi^0} = \left( \begin{array}{ccccccc}
0 & \sfrac{y\sin\beta}{\sqrt{2}} h \, {\bf 1} & 0 & 0  & & & 
\sfrac{y}{\sqrt{2}} s \delta^3_a \\
\sfrac{y\sin\beta}{\sqrt{2}} h \, {\bf 1}  &0& X &
 -\sfrac23 g' s \delta^3_a
& &  \\
0 & X^T & {\cal M}_3 & 0 & & & \\
0 & -\sfrac23 g' s \delta^3_a & 0 & \ppp m_1 & \ppp 0 & -\eta_1' &
\ppp\eta_2' \\
 & & & \ppp 0 & \ppp m_2 & \ppp\eta_1 &-\eta_2 \\
 & & & -\eta_1' & \ppp\eta_1 & \ppp 0 
& -\mu \\
\sfrac{y}{\sqrt{2}} s \delta^3_a  & & & 
\ppp\eta_2' & -\eta_2 & -\mu & \ppp 0 \\
	 \end{array} \right)\, 
\label{horrid2}
\eeq
where $\bf 1$ is the unit matrix in color space, $\delta_a^3$ projects
onto the color breaking direction, and the 
submatrices for the gluinos and gluino-$t_R$ mixing are given by
\beq
	{\cal M}_3 = m_3 \left( \begin{array}{ccccc}
	0 & 1 &   &   & \\
	1 & 0 &   &   & \\
	  &   & 0 & 1 & \\
          &   & 1 & 0 & \\
	  &   &   &   & 1 
	 \end{array} \right)\, ;\quad
	X = \sfrac{1}{\sqrt{2}}\, g_s s \left( \begin{array}{ccccc}
	1 & 0 & 0 & 0 & 0 \\
	0 & 0 & 1 & 0 & 0 \\
	0 & 0 & 0 & 0 & \sqrt{2/3} 
	 \end{array} \right)\, .
\eeq

Finally, let us mention the scalars which remain 
unmixed: the 3 Higgs and 5 right stop Goldstone bosons, with
respective masses (in Landau gauge, used throughout)
\beqa
m^2_{\chi_h} &=& \lambda_h (h^2 - v^2) + \sfrac12\, \lambda_y s^2,\\
m^2_{\chi_s} &=& \lambda_s s^2 - \mu_s^2 + \sfrac12\, \lambda_y h^2.
\eeqa
Also because we work in Landau gauge, the ghosts are massless and do not
contribute to the one loop effective potential.
This completes the list of all particles appearing in the sums for
the one-loop potential.

In computing the above masses, we evaluate the gauge, Yukawa, and
scalar couplings at a common renormalization point $\overline{\mu}$, in
the six quark plus right squark scheme, so the gluino, Higgsino, and
heavy squarks are treated as integrated out.  The scalar couplings are
then the same as the ones appearing in the tree potential.  The value
of $\overline{\mu}$ is a parameter of our effective potential.  The
$\overline{\mu}$ dependence should formally be a two loop effect.
However this does not guarantee it to be as small as might be
expected.  The thermal contributions are formally a one loop effect,
but because the theory has scalar masses which are unprotected from
large radiative corrections (in the absence of SUSY, which thermal
effects break), the thermal potential can correct mass parameters at
order 1.  The $\overline{\mu}$ dependence of the thermal part is only
down by one loop, so $c_h$ and $c_s$ depend on $\overline{\mu}$ at one
loop.  Varying $\overline{\mu}$ gives a good indication of the
sensitivity of our results to two loop thermal effects, in particular
the two loop effects which fix the one loop renormalization scale of
$c_s$ and $c_h$.

\subsection{Thermal masses}

To complete our construction of the effective potential, we need to
determine the thermal masses $\Pi_i(T)$ which are resummed in $V_{1,\rm
therm}(m^2_i)$ by replacing $m^2_i$ with $m^2_i+\Pi_i$. In the
high-temperature limit, these thermal self-energies, of the form $\Pi_i
= c_i T^2$, have all been computed in ref.\ \cite{ComEsp}, which shows
the separate contribution to each $c_i$ coming from every possible
particle in the spectrum of the MSSM.  One should omit the
contributions from any states that are much heavier than the
temperature.  For those which may be on the borderline for thermal
decoupling, say particle $j$, we can flag their contributions by
multiplying them with a coefficient $\theta_j$, in the notation of
\cite{ComEsp}.

Thus, with the spectrum we have assumed, the thermal mass coefficients
for the longitudinal components of the U(1), SU(2) and SU(3) gauge
bosons ($B$, $W$, $A$) are, respectively,
\beqa
	c_{\sss B} &=& \frac{g'^2}{18} ( 41 + 3\theta_{\tilde h} )\\
	c_{\sss W} &=& \frac{g^2}{6} ( 11 + 2\theta_{\sss \widetilde W} 
	+ \theta_{\tilde h} )\\
	c_{\sss A} &=& \frac{g_s^2}{6} (13 + 3\theta_{\tilde g}),
\eeqa
while the transverse components remain massless at this order in the
couplings.  The $\theta_j$ functions interpolate between
1 and 0 as the mass of particle $j$ goes from zero to infinity.
The expression for a fermion is the bracketed part of 
Eq.\ (\ref{therm_mass}), and the expression for bosons is similar but
with the replacements $\exp(x)+1\to\exp(x)-1$ and  $12 \to 6$.
We evaluate the Debye masses at $h=s=0$.

However for the Higgs bosons and stops, there is an added complication;
the Higgs and stop fields themselves give a contribution to the thermal
masses, which are thermal mass dependent.  We self-consistently
determine $\Pi_h$ and $\Pi_s$ so that they really represent the
curvature of $V_{1,\rm therm}$ at the origin of field space, by defining 
\beqa
	\Pi_h &=& {\partial^2 \over \partial h^2} \left. 
	V_{1,\rm therm}(m^2_i(h,s) + \Pi_i)\right|_{h=s=0} \, , \\
	\Pi_s &=&  {\partial^2 \over \partial s^2} \left.
	V_{1,\rm therm}(m^2_i(h,s) + \Pi_i)\right|_{h=s=0} \, .
\eeqa
These relations are recursive, so they cannot be solved analytically,
but they converge very quickly on iteration.
The same thermal mass values also apply to the respective Goldstone
modes of the Higgs boson and the stop.

The fermions' behavior is infrared-safe and there is no need to perform
any mass resummation for them.

\subsection{Two-loop effects}
\def\figeight{\mathrel{\hbox{\rlap{\hbox{\lower2pt\hbox{$\circ$}}}\hbox{\raise2.5pt\hbox{$\circ$}}}}}

We have also considered the effect of including finite-temperature
two-loop contributions to the effective potential.  There are many such
diagrams, which either have the topology of a figure eight
($\figeight$) or the setting sun ($\ominus$).  In the latter, the
trilinear vertex could come from a quartic coupling expanded around the
arbitrary background Higgs or squark field VEV's, or it could represent
cubic couplings involving gauge bosons or gauge bosons and matter
fields.

We have simplified the computation of the two-loop diagrams by ignoring
the $g'$ coupling, which eliminates mixing between the gluon $A_8$ and
the $B$ and $W_3$ gauge bosons.  We also work only to leading order in
the high temperature expansion and treat only degrees of freedom which
are light and therefore influence the strengths of the phase
transitions out of the symmetric phase.  This is appropriate if our
main goal is to understand these transitions more accurately, and it
allows us to use the expressions derived in \cite{John}.  However this
procedure makes two errors: it does not completely account for two-loop
corrections to $c_s$ and $c_h$, and it becomes less accurate at lower
temperatures and larger field values, where the CCB to EW transition
may occur.  We can compensate for the first problem by seeing how large
a correction to $c_s$ must be by artificially inserting a shift $\delta
c_s$ ``by hand,'' but the second error is more problematic.  However,
in this regime the two-loop effects are substantially smaller than the
one-loop effects, which we are treating carefully; and in any case the
form of the two-loop contributions are not known beyond leading order
in the high temperature expansion so it is difficult for us to do
better.

Because of these limitations in the two-loop formulas, we consider their
effects to be indicative of what one might expect from a more careful
treatment, but not necessarily quantitatively accurate.  The good news
is that the two-loop effects tend to make the tunneling from CCB to
electroweak phases more difficult, thus strengthening our conclusions.
It seems likely that the result of a more accurate two-loop treatment would
be somewhere in between those of the one-loop potential and the
high-$T$ expansion of the two-loop potential.

\section{Bubble nucleation from CCB phase}
\label{detail_sec}

In this section we will first discuss how to compute bubble nucleation
rates.  Then we discuss the two problems we need to apply it to:
the problem of getting {\em into} the CCB phase without getting into the
EW phase first; and the problem of getting out of the CCB phase to the
EW phase.

\subsection{Nucleation rates}

To compute the rate of bubble nucleation  at one loop, one should first
find the saddle point of the approximate effective action
\beq
S = \int_0^{1/T} d\tau \int d^3 x \left[ \frac{1}{2} \left( 
	(\partial_\tau h)^2 + (\partial_i h)^2 +
	(\partial_\tau s)^2 + (\partial_i s)^2 \right) + 
	V_{\rm 1 \; loop \, , \; thermal}(h,s) \right] \, .
\eeq
After finding
the saddle point, one should next compute the one-loop fluctuation
determinant about this saddle point, subtracting out those effects
already included by using the one-loop effective potential.  By
incorporating one-loop, thermal effects into the effective potential,
and then subtracting them off from the fluctuation determinant, one
automatically includes the dominant effects in the saddle action.  The
fluctuation determinant then serves to fix the wave function
normalization and account for small additional $O(\alpha_s)$
corrections which can be roughly thought of as higher derivative
corrections.

We will make one simplification and one approximation.  The
simplification is that, at reasonably large temperatures, the saddle
solution does not vary in the (Euclidean) time direction, so the $\tau$
integral can be performed immediately, $\int_0^{1/T} d\tau = 1/T$, and
$\exp(-S)$ becomes $\exp(-E/T)$.  This simplification is strictly
correct down to a temperature $T \sim \omega_- / 2 \pi$, with $\omega_-$
the unstable frequency of the saddlepoint.  Parametrically $\omega_- \sim
m_W$ but numerically it is smaller, and the thermal treatment works down
to $T < 5$  GeV in our case.  We can probe its breakdown by computing
the vacuum action, in which $\int_0^{1/T} d\tau$ is approximated by
$\int_{-\infty}^{\infty} d\tau$.  We find in practice that the tunneling
rate has always peaked at temperatures well above the temperature where
the thermal treatment breaks down, so we are not missing anything by
making this simplification.

The approximation we make is that, rather than computing the full
fluctuation determinant, we approximate its effect by the use of the
one loop thermal effective potential and by a choice of wave function
for the $h$ and $s$ fields such that the curvatures of the potential at
the EW minimum are the physical masses.  This leaves an $O(\alpha_s)$
error in the determined exponent, from the field dependence of the wave
function and from higher derivative corrections.  The error is small
when the phase transition is strong, which indeed is the case, as we
will discuss below.  Our procedure also eliminates renormalization
point dependence at the one loop level.

We use the full one loop effective potential including all
SUSY partners which give vacuum radiative corrections involving strong
or Yukawa couplings.  We do not use a high temperature expansion
or dimensional reduction.  This
avoidance of the high $T$ expansion is appropriate because nucleation
from the CCB to the EW minimum is most likely at a
temperature well below the CCB phase transition temperature, as will be
shown; hence, the field condensates are large and the temperature is
moderate where the nucleation is most likely to occur.  Since the high
$T$ approximation is an expansion in $yh/2\pi T$ or $g_s s/2\pi T$, its
convergence is not very good in the relevant regime.  In contrast,
the loop counting parameter for perturbation theory is $g_s^2 T /
4 \pi g_s s$ or $y^2 T / 4 \pi y h$, which is small.  Two-loop effects
are therefore not expected to be very large.  Because the two-loop
contributions to the effective potential have been calculated only at
leading order in the high $T$ expansion, including them might not
really improve the accuracy of the calculation of the CCB to EW
tunneling action.  On the other hand, the transition from the symmetric
to the CCB phase occurs at a higher temperature, so neglect of the two
loop thermal effects may not be such a good approximation there:  we
make an error in the determination of the phase transition temperature
where the $s$ condensate forms.  But what really matters is the error
in the temperature {\it difference} between the CCB and EW phase
transition temperatures, and we will study how important such an error
is in due course.

Superficially, it may seem that we have made contradictory
approximations:  the effective potential should not rely upon a large
$T$ expansion, while the bubble nucleation treatment can do so.  But
the two statements are actually compatible; the high temperature
approximation for bubble nucleation has a much wider range of validity
than the high $T$ expansion of the effective potential.  This is
because the thermal tunneling treatment depends on $\omega_-$, which
though parametrically of order $m_W$ is numerically  smaller.  Also and
more importantly, the thermal tunneling treatment remains strictly
valid until $T \sim \omega_-/2 \pi$, while the high $T$ expansion
ceases to converge at $T \sim m_t / \pi$ but starts getting large high
order corrections well before then.

If we wanted to perform a complete two loop calculation we would need
not only the one loop fluctuation determinant, but also the two loop
analog.  There are serious technical obstacles to setting up such a
calculation, and we are not aware of any work in the literature which
performs such a calculation for any nontrivial saddle point in a field
theory.  It is an assumption, perhaps justified, that the most important
two loop effects can be incorporated by using the two loop effective
potential.  This is what we do to compare the one and two loop tunneling
rates; the ``two loop'' results discussed below still do not include
even the one loop fluctuation determinant.

\subsection{Getting into the CCB phase:  choice of $\mu_s^2$}

As pointed out in Section \ref{general_sec}, we need a large enough value of
$\mu^2_s$ (the negative stop mass term) to get into the CCB phase
before the electroweak phase transition; but too large a value
prohibits nucleation from the CCB to the EW phase.  So what value of
$\mu^2_s$ should we use?  Since we are trying to see if nucleation from
the CCB to the EW minimum is ever possible, we should use the lowest
permissible value, that is the lowest value for which the symmetric to
CCB transition happens before the symmetric to EW transition can occur.

At this point it is important to distinguish
between the critical temperature $T_c$ and the nucleation temperature
$T_{\rm nuc}$ for a phase transition.  The critical temperature for the
symmetric to CCB phase transition, $T_{c1}$, is the temperature where
the free energies of the CCB phase and of the symmetric phase are equal.
However, the phase transition does not begin until the CCB phase is
favorable enough so that copious bubbles of the CCB phase form.
Roughly, this occurs when the tunneling action of a critical bubble of
the CCB phase is small enough to put one bubble in each Hubble volume
in one Hubble time, $E_{\rm crit}/T \simeq 4 \log(T/H)$, 
with $H$ the Hubble constant.  At the electroweak epoch, $4 \log(T/H)
\simeq 145$.

It is convenient to define, not a nucleation temperature, but a
nucleation temperature range, where the upper edge of the range is the
temperature where there will be one bubble nucleation per horizon volume
and the lower edge is where the phase transition will complete and the
old phase will be completely eaten up.  These differ because the phase
transition takes much less than one Hubble time to occur.  If we define
$f = T dE_{\rm crit}/dT$, then $(1/f) \sim 10^{-4}$ characterizes what
fraction of a Hubble time it takes for the nucleation rate to change
significantly.  The upper edge of the nucleation temperature range
occurs when $E/T = 4 \log(T/H) - \log(f) \simeq 140$.  The single power
of $1/f$ is because there is much less than a Hubble time in which to
put one bubble per horizon volume.  The lower edge of the nucleation
temperature range, where the phase transition completes, is where
$E/T = 4 \log(T/H) - 4 \log(f) \simeq 110$.  The four powers of $1/f$ are
because the bubbles must nucleate close enough together to merge in
$1/f$ of a Hubble time; so there is one power of $1/f$ for each space
dimension and for time.

\begin{figure}[t]
\centerline{\epsfxsize=4in\epsfbox{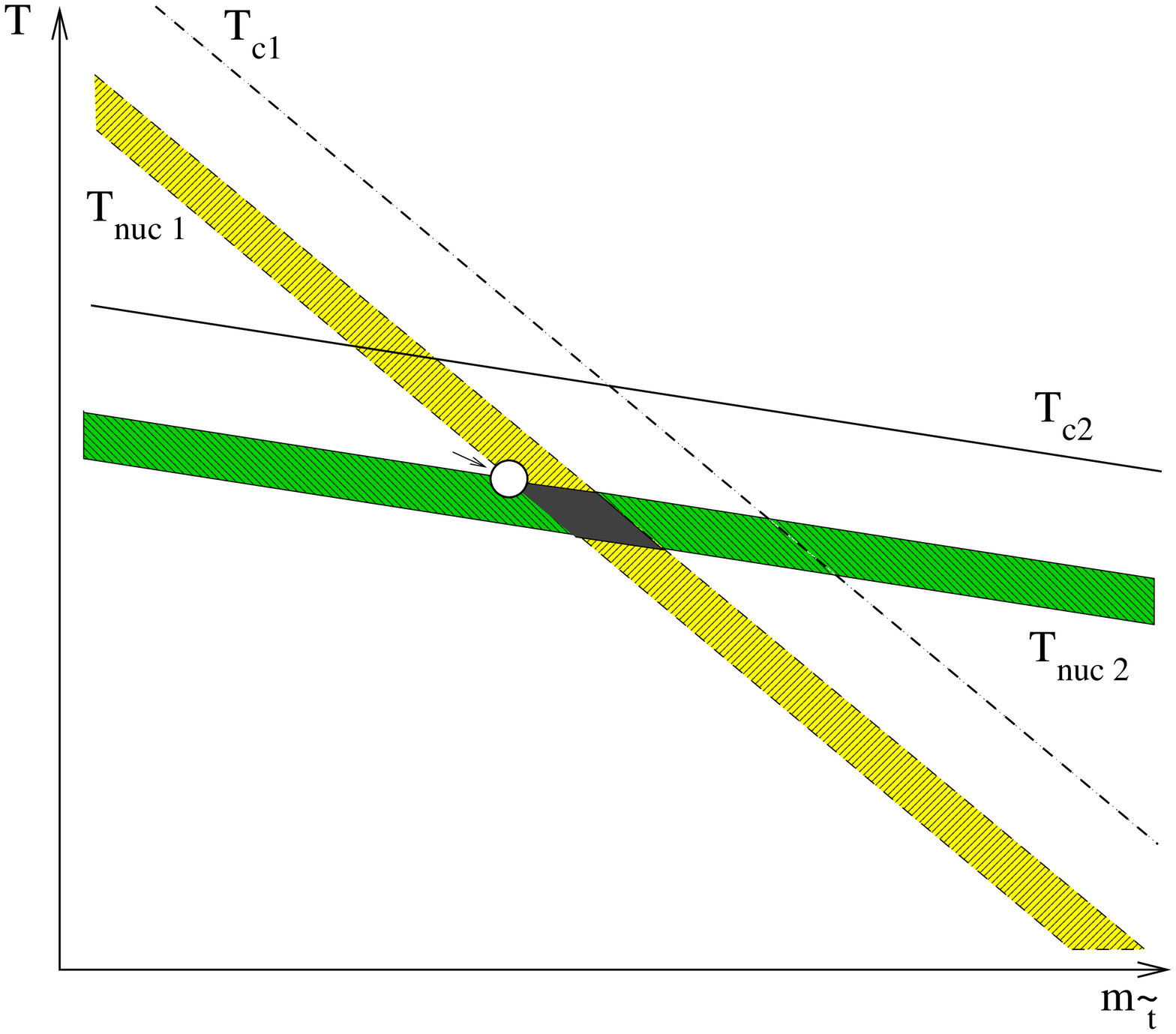}} 
\caption{\label{tempfig} Schematic dependence of critical and
bubble nucleation
temperatures for the two competing phase transitions (symmetric to CCB and 
symmetric to electroweak) as a function of squark mass.}
\end{figure}

The criterion for the symmetric to CCB transition to occur first is that
the lower edge of the symmetric to CCB nucleation band be at a higher
temperature than the lower edge of the symmetric to EW nucleation
temperature band.  That is, the symmetric to CCB transition must
complete before one electroweak bubble per horizon nucleates out of the
symmetric phase.
We illustrate this in Figure \ref{tempfig}, which shows
qualitatively how the two critical temperatures, $T_{ci}$, and the
corresponding bubble nucleation temperatures, $T_{{\rm nuc}\ i}$,
depend on the right top squark mass $m_{\tilde t}$.  The temperature
for the transition to color breaking (1) depends much more strongly on
$m_{\tilde t}$ than that for the electroweak transition (2).  
The open circle in the figure marks the region where the symmetric to
CCB transition completes just before nucleation of EW bubbles; it is the
optimal point.  This choice yields the most shallow
possible CCB minimum and thus the greatest probability of being able to
make the subsequent transition from CCB to EW phases.  The position of
the circle illustrates how we choose $m_{\tilde t}$ once the other
parameters of the MSSM have been specified.

What if we pushed $m_{\tilde{t}}$ a little higher?  Then the universe
would pass through the diamond in Figure \ref{tempfig}, where the
nucleation temperature bands overlap.  In this case several bubbles of
EW phase would nucleate per Hubble volume before the CCB transition 
completed.  If the CCB minimum is deeper at the double nucleation
temperature, these bubbles would be absorbed by the CCB phase.  But
if, as may be the case, the electroweak minimum were the deeper one
already at this temperature, then these EW bubbles could continue to
expand and eat up the CCB phase.  In this case we can get the
phenomenology of EW bubbles expanding into a CCB phase, without any CCB
to EW 
bubble nucleations ever occurring.  However, this only happens for a
very narrow range of values for $\mu^2_s$, and it also depends on the EW
minimum being the deeper one, which is not always the case.  This
scenario is cosmologically viable and would be quite interesting, but it
is highly fine tuned.  We will not address it further since the question
we want to answer is whether we can get into our EW vacuum after an
epoch in which {\em all} of space is in the CCB phase.

\subsection{CCB to EW transition}

Next, let us establish the criterion for judging whether bubble
nucleations are efficient enough to get us out of the CCB phase.
A rough, conservative requirement is that the nucleation barrier has to
be low enough to allow one 
critical bubble of EW phase per horizon volume per Hubble time, $E/T
\simeq 4 \ln(T/H)$.  As long as the universe is dominated by the energy
density of the plasma, $H \sim T^2 / m_{\rm pl}$.  However, at low
temperatures the energy density is dominated by the vacuum energy of
the CCB phase\footnote{unless the CCB phase vacuum energy is negative,
but then tunneling out of it would be impossible.}, which is of order
$m_W^4$.  Thus the Hubble constant never gets parametrically smaller
than $m_W^2 / m_{\rm pl}$.  If we remain in the CCB vacuum when its
vacuum energy becomes dominant then the universe begins to inflate.  If
the nucleation rate continues to be too small at this point, the model
is unacceptable for the same reason that old inflation is
\cite{old_inflation}.  Hence a generous criterion is that CCB to EW
nucleation never takes place if $E/T$ remains greater than\footnote{We
are also assuming that there are no big surprises waiting for us in the
fluctuation determinant; but this seems likely, see
\protect{\cite{Baacke_bubble}}.} $4 \ln(m_{\rm pl} / m_W) + 4 \ln 10
\simeq 170$, where the extra term $4 \ln 10$ is a cushion to insure
that our conclusions will be robust.

It is easy to see that nucleation from the CCB to the EW phase can
never occur {\it immediately} after the CCB phase transition.  We
already arranged for the symmetric to EW transition to be slower than
the symmetric to CCB transition; the CCB to EW transition will be even
slower for two reasons:
\begin{enumerate}
\item The separation in field space between EW
and CCB minima is larger than that between symmetric and EW minima;
\item the CCB minimum is necessarily deeper than the symmetric one at
$T_{\rm nuc1}$, so the potential difference between the CCB and EW
minima is smaller than between symmetric and EW.  
\end{enumerate}
Both of these factors make the CCB $\to$ EW transition slower than the
symmetric $\to$ EW one.  
Therefore if the temperature $T_{\rm nuc 3}$ exists, where the CCB
phase nucleates copious bubbles of EW phase, it must be considerably
below $T_{\rm nuc 1}$.  The CCB and EW minima become ever deeper and
the squark and Higgs condensate values become larger as $T$ falls, so
the separation of the minima becomes larger.  This is why the high $T$
expansion is not necessarily reliable at $T_{\rm nuc 3}$, whereas
perturbation theory is {\it more} reliable than at the previous phase
transition.

We can summarize our procedure as follows.  The vacuum theory retains
one free parameter we have not yet fixed, $\tilde{A}$ the mixing
parameter.  We examine values from zero mixing up to the largest
$\tilde{A}$ that is compatible with the experimental lower limit on the
stop mass.  At each value we find the $m_Q$ which gives $m_H({\rm
physical}) = 95$  GeV and the smallest $\mu^2_s$ for which the CCB
transition happens before the EW one.  Then we compute the tunneling
action from the CCB to the EW minimum for a range of temperatures
between $T_{\rm nuc1}$ and  5  GeV, as well as the vacuum ($T=0$)
tunneling action.  The bubble action is determined using a new and very
efficient algorithm presented in Appendix \ref{AppendixA}.  We confirm
that tunneling is always inefficient at $T_{\rm nuc1}$; its rate
usually peaks at some intermediate temperature, roughly $(2/3) T_{\rm
nuc 1}$.  We also confirm that vacuum tunneling is always extremely
inefficient, so much so that typically the thermal tunneling treatment
gives the larger (hence correct) value for the rate down to
temperatures as low as 3  GeV.

\section{Results and Conclusions}
\label{Conclusion_sec}

In this section we present our results for the energy $E$ of the bubble
solutions which interpolate between the CCB and EW vacua, and show that
$E/T$ is always larger than the value needed for the phase transition
to complete.  We will then discuss what kind of new physics might be
able to change this conclusion, and the constraints on the MSSM which
our analysis implies.

\subsection{Results}

Using the one loop effective potential with a renormalization point
$\overline{\mu}=150{\rm\  GeV}$ intermediate between the top and right
stop masses, and at zero squark mixing $\tilde A =0$, we find that the
minimum value of $E/T$ over temperatures is $1340$, giving a tunneling
rate per unit volume of order $T^4 \exp(-1340)$, which is drastically
smaller than the required value of $T^4 \exp(-170)$.  The physical stop
mass in this zero-mixing case is $126 {\rm\  GeV}$, which is lower than
might be expected because of the large downward radiative corrections
to $\lambda_y$.

\begin{figure}[t]
\centerline{\mbox{\epsfxsize=3in\epsfbox{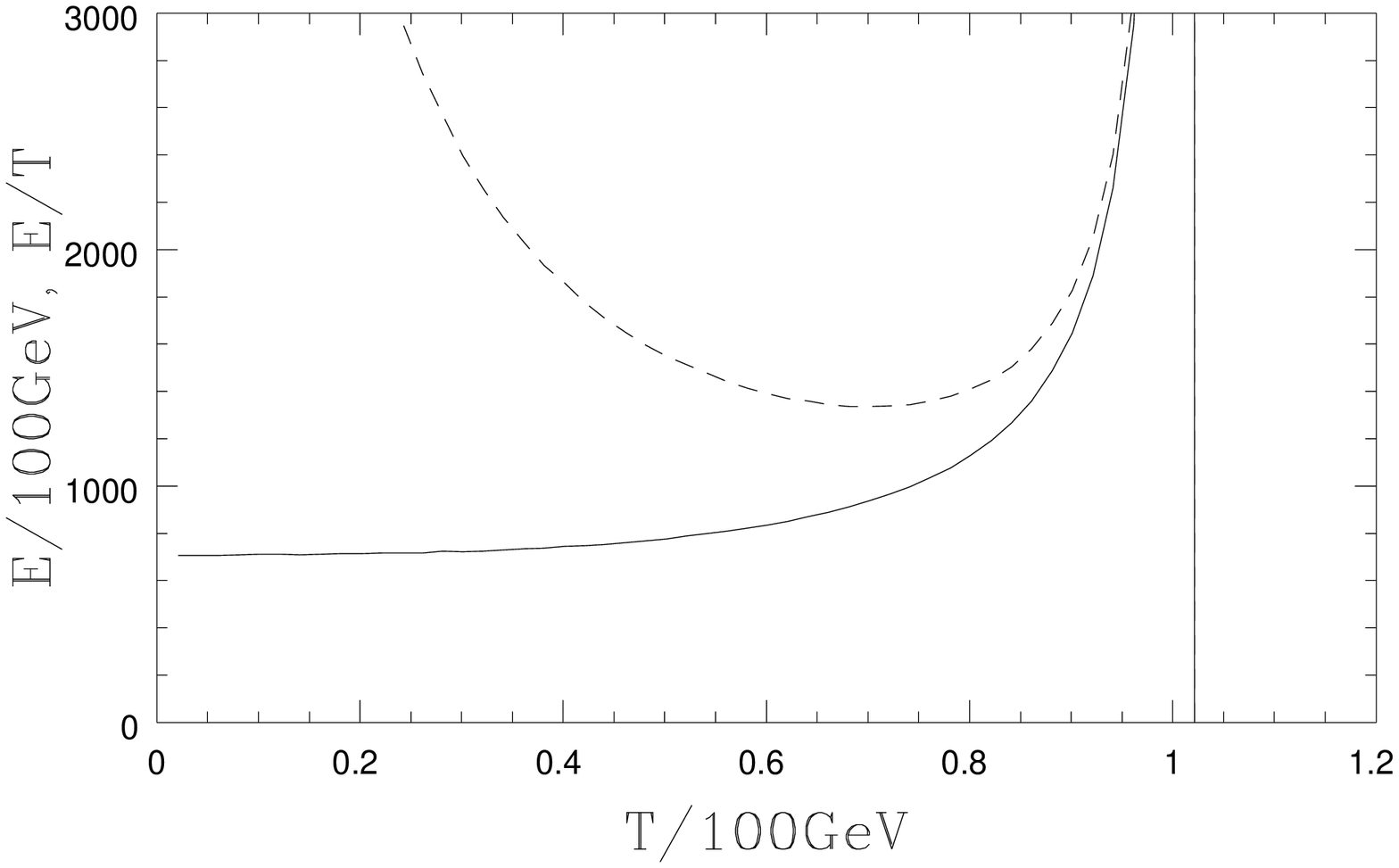}} \hspace{0.4in}
\mbox{\epsfxsize=3in\epsfbox{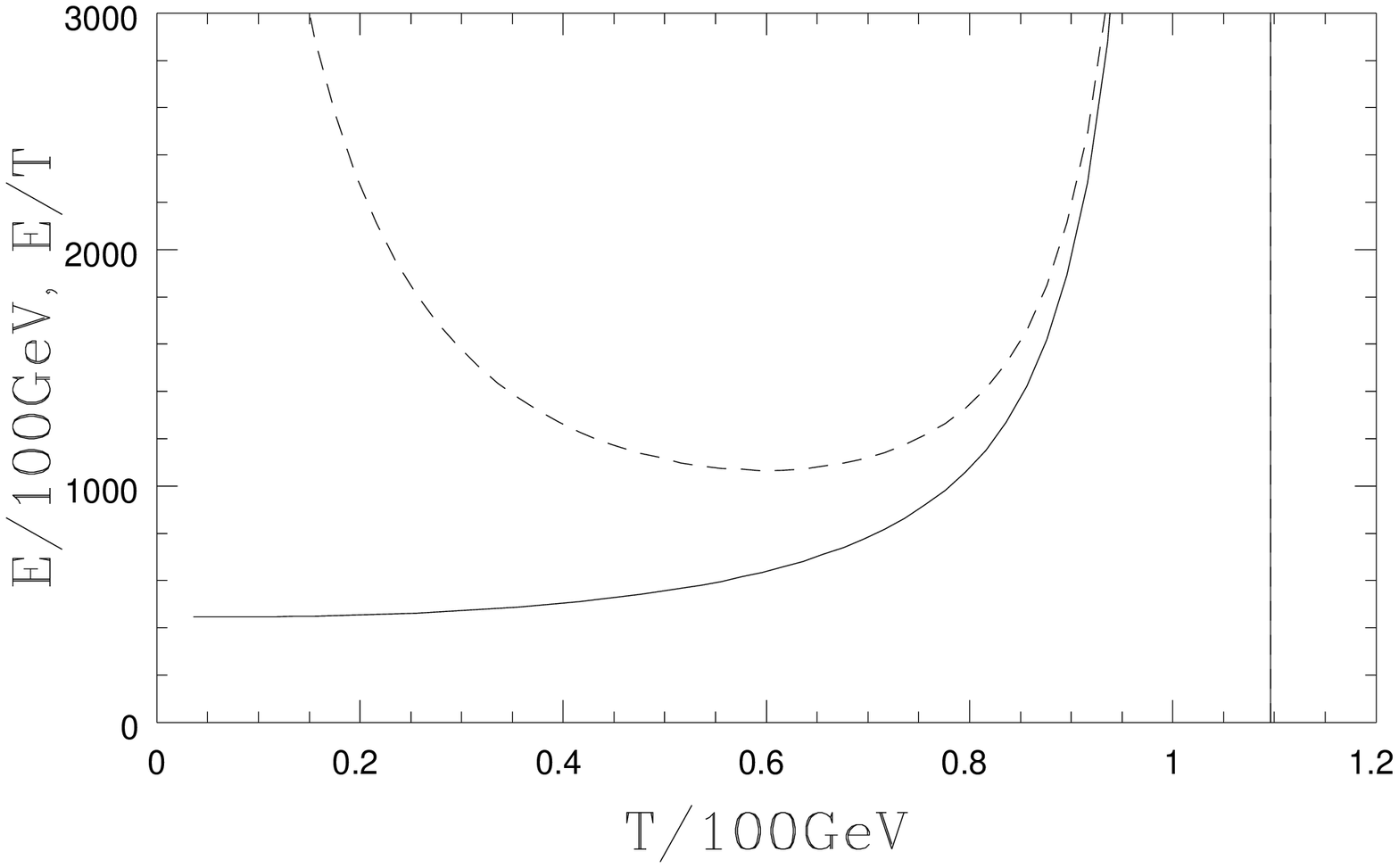}}}
\caption{\label{Efig} $E$ in units of $100{\rm\  GeV}$ (solid line) 
and $E/T$ (dotted line)
as a function of temperature, for the critical bubble mediating
the CCB to EW phase transition.  The left-hand figure is for the case of
no mixing and the right-hand figure is maximal mixing, so the right stop
mass is $90{\rm\  GeV}$.  The vertical bar is $T_{\rm nuc1}$.}
\end{figure}

\begin{figure}[p]
\centerline{\mbox{\epsfxsize=3in\epsfbox{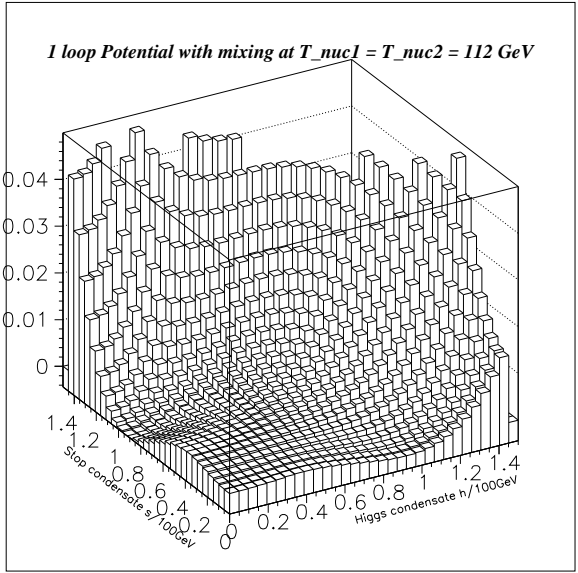}} \hspace{0.4in}
\mbox{\epsfxsize=3in\epsfbox{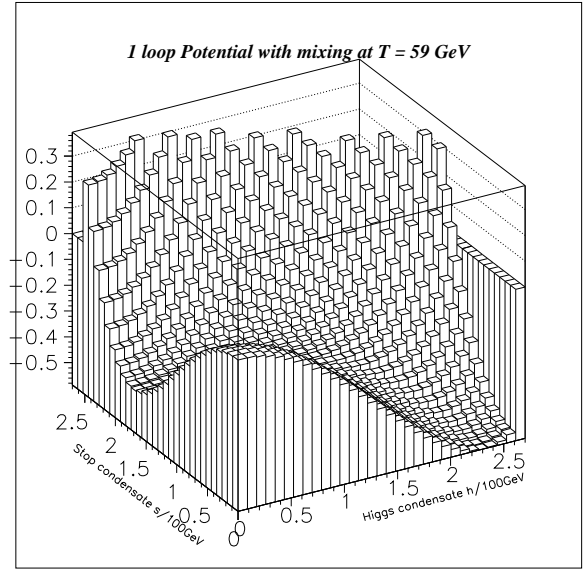}}}
\caption{\label{potpic1} The potential one loop potential, with squark
mixing, at $T_{\rm nuc1}$ (left) and at $T$ which minimizes $E/T$
(right).  Although the CCB minimum near vacuum becomes quite
shallow, it is still not shallow enough to allow efficient nucleation.
Note scales; the vertical scales are in units of $(100 {\rm\  GeV})^4$.}
\end{figure}

\begin{figure}[p]
\centerline{\mbox{\epsfxsize=3in\epsfbox{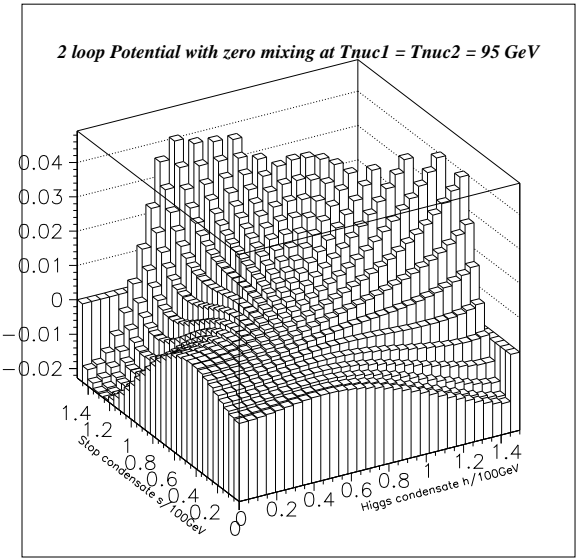}} \hspace{0.4in}
\mbox{\epsfxsize=3in\epsfbox{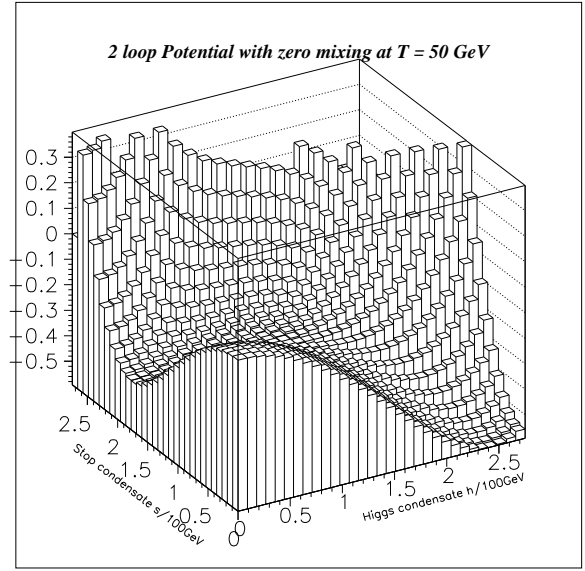}}}
\caption{\label{potpic2} This figure is the same as Figure
{\protect{\ref{potpic1}}} except that there is maximal allowed mixing
and two loop effects are included in the effective potential.
The CCB transition is stronger, so the CCB minimum is deeper and harder
to get out of.}
\end{figure}

Mixing between the left and right stops helps but only weakly; mixing
maximally so that the stop mass saturates its experimental bound
reduces $E/T$ to $990$, which is still far too large to allow the
phase transition to complete.  The dependence of the tunneling energy
on temperature is shown for each of these cases in Figure \ref{Efig}.
The energy of the critical bubble is large at high temperatures and
falls monotonically as the temperature is reduced.  Likewise the
tunneling action is large immediately after the symmetric to CCB
transition; in fact, at zero mixing, there is a range of temperatures
immediately below $T_{\rm nuc1}$ for which tunneling to the EW minimum
is kinematically forbidden.  We illustrate the potential as a function
of Higgs and squark fields, both at $T_{\rm nuc1}$ and the temperature
where $E/T$ is minimized, in Figures \ref{potpic1} and \ref{potpic2}.

\begin{figure}[t]
\centerline{\epsfxsize=5in\epsfbox{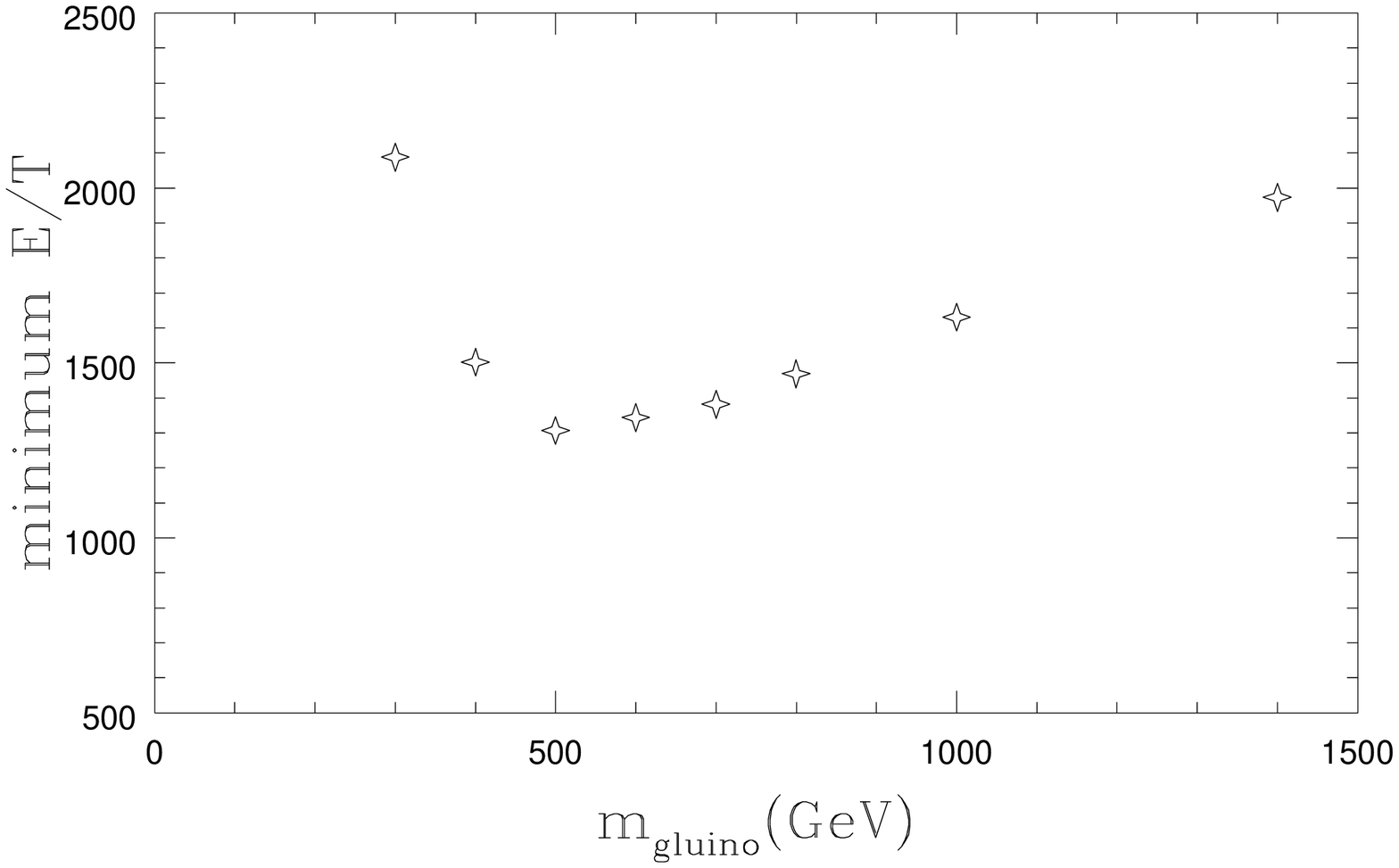}}
\caption{\label{mgdep} Minimum value of the bubble action,
$E/T$, as a function of the gluino mass, using the one loop potential
and at zero mixing.}
\end{figure}

To verify the arguments of section \ref{rmpsubsect}, we have also
checked that our choice of particle masses is optimal.  In particular,
if the Higgsino or gluino are allowed to be lighter it makes the
transition much harder, and if the gluino is heavier the minimum $E/T$
also rises quickly because of the large correction to $\lambda_s$.
This behavior is shown in Figure \ref{mgdep}.  Making the Higgsino
heavier has a less dramatic effect, but it is also unfavorable to
tunneling.  To see whether the assumptions about the other particle
masses are important, we have pushed the superheavy squark mass scale
all the way to $10^{10}{\rm\  GeV}$, and the left stop mass as high as
$20{\rm\ TeV}$, obtaining a minimum value of $E/T = 1010$ in the
zero-mixing limit.  This demonstrates that the choice of masses for the
very heavy scale particles have no qualitative effect on our
conclusions.

We have also checked the robustness of our results with respect to
changing the renormalization point.  The primary effect of varying
$\overline\mu$ is to change the thermal contributions to the effective
potential, as we have discussed.  Setting $\overline{\mu}=90{\rm\  GeV}$
raises the minimum $E/T$ without mixing to 1490; choosing
$\overline{\mu}=500{\rm\  GeV}$ lowers $E/T$ to 970 at zero mixing,
or 840 at maximal mixing.  All of these values are still far from that
needed for bubble nucleation.  Varying the renormalization point
roughly accounts for the uncertainty in $c_s$ and $c_h$ from two loop
effects.  The results from the recent paper by Losada
\cite{Losada_latest} show that the best value for the
renormalization point is a few times $T$, which is within the range we
check here; however we were not able to use the explicit expressions
from that paper because it makes different assumptions about what
degrees of freedom are heavy.   It also uses the high temperature
approximation, which as we have stressed is not entirely reliable
in the present context.

Another important check is to see how two loop thermal effects change
our answers.  It has already been observed in previous work that they
strengthen the phase transition from the symmetric to the CCB phase
\cite{John}.
This makes getting out of the CCB phase much harder, both because it
increases the required value of $\mu^2_s$, and because it makes the CCB
minimum deep already at a higher temperature.  As a result, we find
that without mixing, the minimum value of $E/T$ increases to
$3000$.  Even adding ``by hand'' a $20\%$ downward contribution to
$c_s$, the action remains too high, with a minimum $E/T$ of 1220.
In fact, getting the minimum $E/T$ down to 170 requires a ``by hand''
reduction to $c_s$ of $45 \%$, which two loop effects beyond our
leading log treatment cannot possibly provide.

Also, mixing no longer helps when the two loop effects are included.
This is because mixing weakens the electroweak transition
substantially, since the strength of the latter is set
mostly by the coupling of the Higgs to the stop, $\lambda_y$; but
mixing has little effect on the CCB transition, since its strength
comes mainly from gluonic diagrams and not from diagrams involving
$\lambda_y$.  The two loop effects enhance the CCB transition, and if
it is very strong and the EW transition is weak, it is more difficult
to get out of the CCB minimum.  We illustrate this in Figure
\ref{potpic2}, which is the same as Figure \ref{potpic1} except that it
is for maximal mixing and including the two loop effects.

We might also ask, how essential are the experimental bounds on the
Higgs and stop masses to our result?  The bound on the Higgs mass turns
out to be inessential; allowing $m_h$ to go down to $65 {\rm\  GeV}$
still gives a minimum $E/T = 660$, using the one loop potential with
mixing, the most favorable combination.  However, the bound on the stop
mass is essential.  If the mixing is large enough, and hence
$\lambda_y$ small enough, then the second inequality in Eq.
(\ref{stab_cond}) will be violated, and the CCB ``minimum'' will
actually be a saddle.  However, at high temperatures there may still be
a CCB minimum.  In this case the universe can go into the CCB minimum
safely, because at some temperature the CCB minimum becomes spinodally
unstable, and nucleation of EW bubbles is guaranteed to be efficient
just above the spinodal temperature.  The required value for the stop
mass is about $60{\rm\ GeV}$ using the one loop potential and about $50
{\rm\ GeV}$ using the two loop potential.

\begin{figure}[t]
\centerline{\epsfxsize=4in\epsfbox{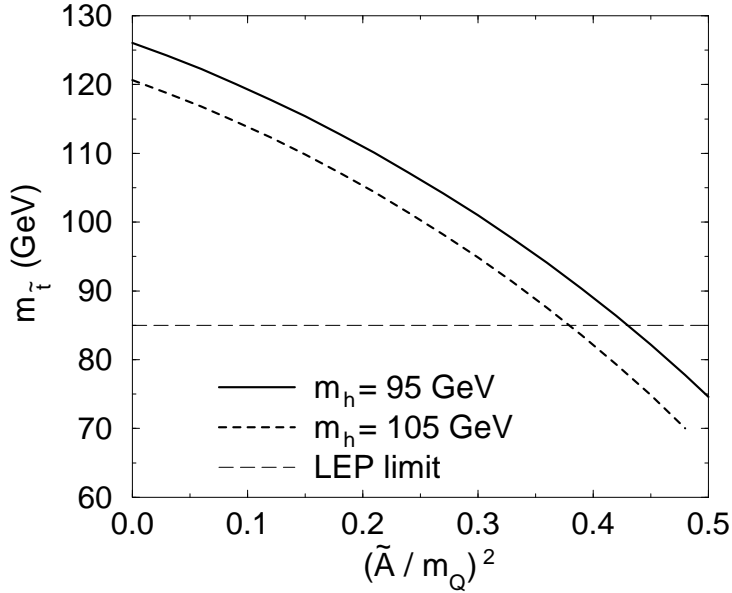}} 
\caption{\label{constraintfig} Our upper limit on the physical stop mass
as a function of $(\tilde A/m_Q)^2$, for two values of the light Higgs boson
mass, using the one-loop effective potential.  The region $m_{\tilde t}
< 85$ GeV is separately excluded by accelerator search limits.}
\end{figure}

The fact that color breaking is ruled out allows us to exclude some 
parameter values in the MSSM, namely those for which the color
breaking nucleation temperature $T_{\rm nuc 1}$ is greater than that of
the electroweak transition, $T_{\rm nuc 2}$.  This condition involves
many unknown quantities, such as $\tan\beta$, the Higgs boson mass $m_h$, the
left stop mass $m_Q$, and the stop mixing parameter $\tilde A$.  We
have illustrated the constraint by fixing $\tan\beta=3.2$, while varying
$\tilde A/m_Q$ and $m_Q$ in such a way as to keep $m_h$ fixed at 95 GeV, and
fixing $\tan\beta=7.5$ and keeping $m_h=105$ GeV.  The
excluded region is a stop mass less than some value which depends on 
$\tilde A/m_Q$, shown in Figure \ref{constraintfig}.  These are relevant
variables because for any value of $m_{\tilde t}$, one can always
avoid the color breaking transition by making the bare stop mass parameter
less negative ({\it i.e.,} letting $\mu_s^2$ be smaller), while increasing
$\tilde A/m_Q$.  Decreasing $\mu_s^2$ increases $m_{\tilde t}$ while 
increasing $\tilde A/m_Q$ does the opposite, so one can keep $m_{\tilde t}$
fixed by adjusting the two.  To get $m_h$ large enough, both $\tilde A$ and
$m_Q$ take values in the TeV.  We find that the limiting curves are
quite insensitive to the gluino mass.

\subsection{Is there a way out?}

The most efficient way of evading our negative
result is to find some new physics that decreases the thermal
contributions to the right-handed stop Debye mass.  Although no such
effects are present within the MSSM, one can imagine loopholes
in extended models, such as those without $R$-parity.  Here we give
just one example.

In the absence of $R$-parity, the superpotential includes the
baryon number violating terms
\beq
	y'_{ijk} \epsilon^{abc} U^a_i\; D^b_j\; D^c_k\, ,
\eeq
involving the right-handed up ($U$) and down ($D$) squark fields of
generation $i,j,k$ and color $a,b,c$, with $y'_{ijk}$ antisymmetric
under $j\leftrightarrow k$.  
It is possible for $y'_{332}$ to be large, if other
$R$-parity violating couplings are sufficiently small, without
violating any experimental constraints.  Associated with the above 
coupling, one anticipates soft SUSY-breaking terms in the potential
of the form
\beq
	y'_{332}\; A'\; \tilde t_R^3\; (\tilde b^1_R\; \tilde s^2_R - 
		\tilde b^2_R\; \tilde s^1_R )\,.
\eeq
When the stop condenses, $\tilde t_R^3 = s$, 
it induces mixing between the bottom
and strange squarks, giving a mass matrix of the form
\beq
	\left( \begin{array}{cc} m^2_{\tilde s} & \pm y'A's \\
				\pm y'A's & m^2_{\tilde b}
	\end{array} \right)\, .
\eeq
Let us consider the situation where there is a hierarchy between the
strange and bottom squark diagonal masses, 
$m^2_{\tilde b} \gg m^2_{\tilde s}$.  The lighter squark gets a 
negative correction
to its mass eigenvalue from the mixing,
\beq
	m^2_{\tilde s} \to m^2_{\tilde s} - 
	{(y'A's)^2\over m^2_{\tilde b} }
\eeq
which makes a negative 
contribution to the stop thermal mass ($c_s T^2$) from the one-loop
finite temperature potential,
\beq
	\delta c_s = - {(y'A')^2\over 6\; m^2_{\tilde b} }\;.
\eeq
Although the heavier squark would make an equal and opposite
Contribution, it is suppressed if $m_{\tilde b}\gg
T$.  The shift $\delta c_s$ could conceivably be large enough to reduce
$c_s$ by the 45\% needed in order to make the CCB to electroweak
transition occur.

Another way of thinking of this is that the trilinear term
has induced a negative quartic coupling between the
strange and stop squarks, analogous to the negative contribution
$\tilde{A}$ made to $\lambda_y$.  A negative coupling between scalars
leads to negative thermal masses, which is the physics of thermal
symmetry non-restoration.
However, for this to work it is essential that there
are very large $R$ parity violating effects involving rather light
squarks.  It is also a little dangerous to induce such a negative
effective quartic coupling; it means that there is a very deep extra
minimum of the potential in which the right stop, right scalar strange
quark, and right scalar bottom quark carry condensates.  It is necessary
that the universe never nucleates into this minimum, and it may be more
problematic to explain the approximate vanishing of the cosmological
constant if ``our'' electroweak minimum is not the global one.

\subsection{Conclusions}

The phenomenology of electroweak bubbles, in which the Higgs field has
a condensate, expanding into a charge and color broken phase where the
right stop has a condensate, is potentially rich, and it could be very
interesting for baryogenesis.  Unfortunately, unless there is new
physics beyond the MSSM, this scenario cannot arise by nucleation of EW
bubbles out of the CCB phase.  We have mentioned $R$-parity violating
interactions as one example of such new physics.  Another could be the
existence of cosmic strings which induce a Higgs field condensate along
their 
cores.  Such defects would act like impurities in a solid state system,
providing sites for the accelerated nucleation of the electroweak
bubbles.  The CCB phase can also appear if both phases nucleate out of
the symmetric one simultaneously, coexisting for a brief period before
the true vacuum state (hopefully electroweak) takes over by squeezing
out the CCB bubbles.  This latter possibility occurs for such a narrow
range of parameter values that we do not consider it to be very
compelling.

Thus in the context of the MSSM and barring any additional physics, we
conclude that cosmology with a stop squark condensate just before the
electroweak phase transition is ruled out.  Under these assumptions we
can exclude MSSM parameter values, such as those shown in Figure
\ref{constraintfig}, which lead to a CCB phase transition temperature
higher than the EW phase transition temperature.

\appendix
\section{Saddle point search algorithms}
\label{AppendixA}

In this section we will describe two algorithms we use for finding
critical bubble actions.  One is a general purpose saddle point finding
algorithm, mentioned also in the appendix of \cite{nostrings}.  The
other is special to finding critical bubbles.  The second algorithm is
highly efficient and to our knowledge it has not appeared previously in
the literature.

\subsection{General saddlepoint finding algorithm}

We want to find a saddle point of a real valued function
$H(q_{\alpha})$, where $q_{\alpha}$ are the set of real degrees of
freedom (or other continuous variables) on which $H$ depends.  In our
particular case, the $q_{\alpha}$ are the values of the Higgs and stop
fields on a discrete set of points representing radii from $r=0$ out to
some $r_{\rm max}$.  The Hamiltonian we want to discretize
is
\beq
H = 4 \pi \int r^2 dr \left[ \frac{1}{2} (\partial_r h)^2 + 
	\frac{1}{2} (\partial_r s)^2 + V(h,s) \right] \, ,
\eeq
where $h$ and $s$ are the Higgs and stop condensates in the real field
normalization and $V(s,h)$ is the thermal effective potential.  
An explicit numerical implementation of $H$ for the
present purposes would be to discretize the radius to integer multiples
of a discrete spacing $\Delta$ and approximate the energy as
\begin{equation}
\frac{H}{4 \pi} = \sum_{i=0}^{i_{\rm max}-1} 
	\frac{i^2 + i}{2} \Delta \left( 
	(h(i+1)-h(i))^2 + (s(i+1)-s(i))^2 \right)
	+ \sum_{i=1}^{i_{\rm max}} i^2 \Delta^3 V(s,h) \, ,
\eeq
This form for the potential is not essential to the algorithm, though;
all we need is for $H$ to depend on a finite number of
coordinates and to possess first derivatives which are easy to evaluate
numerically.  

If we were looking for a minimum of $H$, we could do so by using the
``gradient descent'' algorithm; pick a starting guess $q_{\alpha}(0)$ for
the fields, evaluate the set of derivatives 
\beq
G_\alpha(0) \equiv c_\alpha 
	\left. \frac{\partial H}{\partial q_{\alpha} } 
	\right|_{q = q(0)} \, ,
\eeq
which comprise the gradient of $H$, and update the fields using
\beq
q_\alpha(1) = q_\alpha(0) - \Delta_\tau c_\alpha G_\alpha \, .
\eeq
Here $\Delta_\tau$ is the quenching step length and must be chosen
small enough to make the algorithm stable, and the coefficients
$c_\alpha$ represent a choice of the metric on the space ${q_\alpha}$,
which should be such that the limiting $\Delta_\tau$ to give
stability is approximately the same for excitations involving any $q_\alpha$;
typically $c_\alpha \sim \partial^2 H / \partial q_\alpha^2$.  Then we
define $q_\alpha(n)$ to be the $n$th iterate of the procedure.  This
algorithm converges to a minimum.

The usual approach in the literature to find a saddle point is to derive
from $H$ equations of motion $E_\alpha = \partial H/\partial q_\alpha$,
and then to define $H' = \sum_\alpha d_\alpha E_\alpha^2$, with
$d_\alpha$ some positive coefficients.  A saddle point of $H$ is a
minimum of $H'$, and one can use gradient descent or any other minimum
seeking algorithm.  However this approach can be inefficient if the
saddle point has a very small unstable frequency, and it is also quite
cumbersome because $H'$ is more complicated than $H$; for instance, if
$H$ contains terms with two derivatives, $H'$ has terms with four.  

We have therefore
devised instead an algorithm which deals directly with $H$,
and converges rapidly to the desired saddle point.  
A single iteration of the procedure requires doing the following:
\begin{enumerate}
\item Perform $N$ steps of the gradient descent algorithm, with step
size $\Delta_\tau$.
\item Perform one step of gradient descent with step size $- N
\Delta_\tau$.  Because of the sign, this is actually a ``gradient
ascent'' step, rather than descent.
\item By examining $G_\alpha$ before and after, optimize the
value of N.
\end{enumerate}
On a ``straight slope,'' this algorithm does nothing, because the
gradient ascent step undoes the gradient descent steps.  However,
when the second derivatives of $H$ do not vanish, $N$ forward steps are
not equivalent to one backward step of $N$ times the length.  This is
because each forward step starts where the last one stopped.  On a
concave surface, gradient descent moves towards a stationary point.  
As the slope becomes
smaller, the size of the gradient descent steps becomes smaller.  The
backward step is then $N$ times as long as the {\em smallest} step, and
the final configuration is closer to the bottom than the starting one.
We illustrate this in Figure \ref{step_fig}.
On the other hand, on a convex surface, gradient descent moves away from
the stationary point, and each step is larger than the previous one.
The backward step is $N$ times as large as the {\em largest} forward
step, and overshoots the starting point.  Unless $N$ is too large and it
overshoots too much, the algorithm again lands closer to the stationary
point.  It is to avoid the problem of overshooting in the case where
$N \Delta_\tau$ is too large that the third step, optimizing
$N$, is necessary.

\begin{figure}[t]
\centerline{\epsfxsize=5in\epsfbox{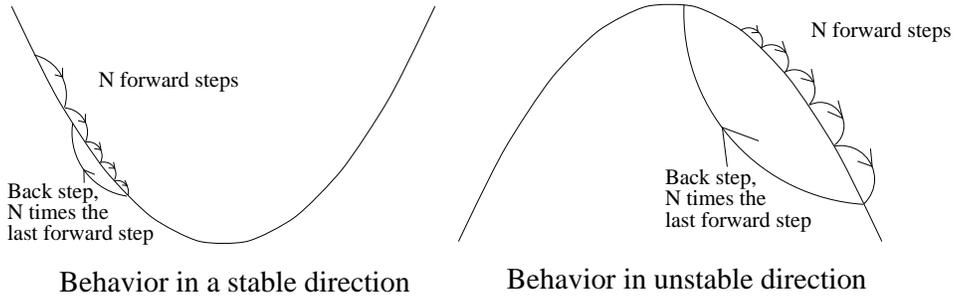}}
\caption{\label{step_fig} Cartoon showing how the saddle seeking algorithm
works.  When an extremum is a minimum, gradient descent steps go towards
it, and the backwards step is smaller than the series of forward steps.
When it is a maximum, the gradient descent moves away, but the backwards
step is larger and overshoots, landing closer to the extremum.}
\end{figure}

Since $H$ is defined in a high dimensional space it is not true
that one or the other of the two circumstances mentioned above pertain.
Close to an extremum, though, $H$ is approximately a quadratic form
in the $q_\alpha$, $H \sim H_{\alpha \beta} \delta q_\alpha 
\delta q_\beta/2$, and the above arguments apply separately for each
eigenvector of $H_{\alpha \beta}$.  More generally, unless $N
\Delta_\tau$ is very large, the algorithm will always go uphill along
directions with negative curvature and downhill along directions with
positive curvature, which will lead it towards a region with smaller
gradients, and hence towards some extremum.

Now we will describe the procedure for optimizing $N$.  First, one notices
that if the departure from the saddle point is predominantly in
convex (stable) directions then we get closer to the minimum fastest
simply by using gradient descent without backward steps.  It is also
easy to tell if this is the case; when it is, $\sum_\alpha G_\alpha
G_\alpha$ diminishes with each forward step.  For this reason, and
because the unstable frequency of a critical bubble is typically lower
than any of the stable frequencies, we will concentrate on the case
where almost all that is left is departure from the saddle in the
unstable direction.  One iteration of the algorithm multiplies the
departure from the saddle in the unstable direction by $(1-x) \exp(x)$,
where $x=N \Delta_\tau \omega_-^2$ and $\omega_-$ is the unstable
frequency of the saddle point.  The algorithm overshoots if $x > 1$ and
it is unstable if $x > 1.278$.  However, we can measure the extent of
overshoot or undershoot by comparing the gradient after an iteration of
the algorithm, $G_\alpha({\rm after})$, with the gradient before,
$G_\alpha({\rm before})$.  Our indicator of whether $N$ is too large is
\beq \frac{ \sum_\alpha G_\alpha({\rm after}) G_\alpha({\rm before}) }
	{\sum_\alpha G_\alpha({\rm before})^2} \, ; \eeq if this is
positive, we can safely increase $N$, and if it is negative we must
reduce $N$.  If there are no remaining excitations in stable directions
then the value of the indicator will be $(1-x) \exp(x)$, which makes it
easy to choose a new value of $N$ which will make $x$ very close to 1.
When $x=1$, the algorithm ``steps back'' just the right distance and
lands on the saddle point.  It is also possible to determine the
unstable frequency from the value of $N \Delta_\tau$ which worked
optimally.

As with any saddle point finding algorithm it is still necessary to feed
in a good starting guess so that the algorithm finds the right extremum
of the action.  Here we have little new to say.  Our approach has been
to define a few-parameter {\it Ansatz} for a path in field space between
the EW and CCB vacua, and to use a shooting algorithm to find the action
for each value of the parameters.  Then we minimize the action over the
parameters in the {\it Ansatz}.  All that is necessary is that the
starting guess not be terribly bad, although in practice the saddle
finding algorithm converges faster if the starting guess is better.

\subsection{Efficient algorithm just for multi-field critical bubbles}

Now we describe a much more efficient algorithm, which is however
special to the problem of determining critical bubble configurations and
actions in theories with more than one field.
The general problem is to find the lowest saddle point of the
Hamiltonian
\beq
H = 4 \pi \int r^2 dr \left( \sum_i \frac{ (\partial_r f_i(r))^2}{2} 
	+ V(f_i(r)) \right) \, ,
\eeq
where $f_i$ represent several fields which may all have condensates, and
the boundary conditions are that the $f_i$ start at $r=0$ near the true
minimum and approach their false vacuum values at large $r$.  Although
we have in mind a numerical implementation involving discretization of
$r$, we use the simpler continuum notation.

The problem reduces to the one field case if we consider a restricted
set of configurations in which the fields always lie along a one dimensional
trajectory
through field space.  That is, we choose a curve in the space of $\{ f_1
, \ldots , f_n \}$, parameterized by a path length $l$, $f_i = f_i(l)$.  
By path length we mean that $l$ is chosen so that
\beq
\sum_i \left( \frac{d f_i(l)}{dl} \right)^2 = 1 \, .
\eeq
Then we require that the fields $f_i(r)$ can be written as $f_i(l(r))$.
This is the same as making all of the fields dependent on the
value of one field.  For this restricted set of configurations, the
Hamiltonian is
\beq
H({\rm restricted}) = 4 \pi \int r^2 dr \left[ \frac{1}{2} \left(
	\frac{dl}{dr} \right)^2 + V(f_i(l(r))) \right] \, .
\eeq
The standard shooting algorithm finds the saddle point on this
restricted class of configurations, and its action is an upper bound for
the true saddlepoint action.  
The ``only'' remaining problem is  to
find the minimum over all choices of paths in field space.

This is where the gradient descent algorithm comes in.  
If our choice of path is imperfect, the shooting algorithm gives a
bubble configuration which is not a true saddle point.  So, lifting the
requirement that the fields lie on any prescribed path in field space,
gradient descent will lead to a lower energy configuration which must,
at least initially, be following a ``better'' path through field space,
meaning one which will give a lower saddle point energy.
This leads to the following algorithm.  First, we choose some
``reasonable'' path through field space.  We evaluate the potential at a
series of points along it and define the potential to be the spline
interpolation of those points.  Then we iterate the following procedure:
\begin{enumerate}
\item Find the saddle point solution for the particular path through
field space by the standard ``shooting'' algorithm;
\item Apply a reasonably short amount of gradient descent cooling to the
resulting configuration, making no requirements that the fields remain
on any trajectory in field space;
\item Use the $f_i(r)$ after the gradient descent to define a new choice
for a path through field space.  In practice we know $f_i$ at a discrete
set of radii $r$, so we take the path to be the series of straight line
segments joining the points $f_i(r_{\rm known})$, 
and the potential to be the spline interpolation of $V(f_i(r))$. 
\end{enumerate}
We illustrate how the iteration converges to the ``right'' path through
field space in Figure \ref{move_through_fieldspace}, which shows a
series of paths in field space from iterations of the above algorithm.
In our case there are only two fields, but the algorithm generalizes
immediately to many fields.

\begin{figure}[t]
\centerline{\epsfxsize=5in\epsfbox{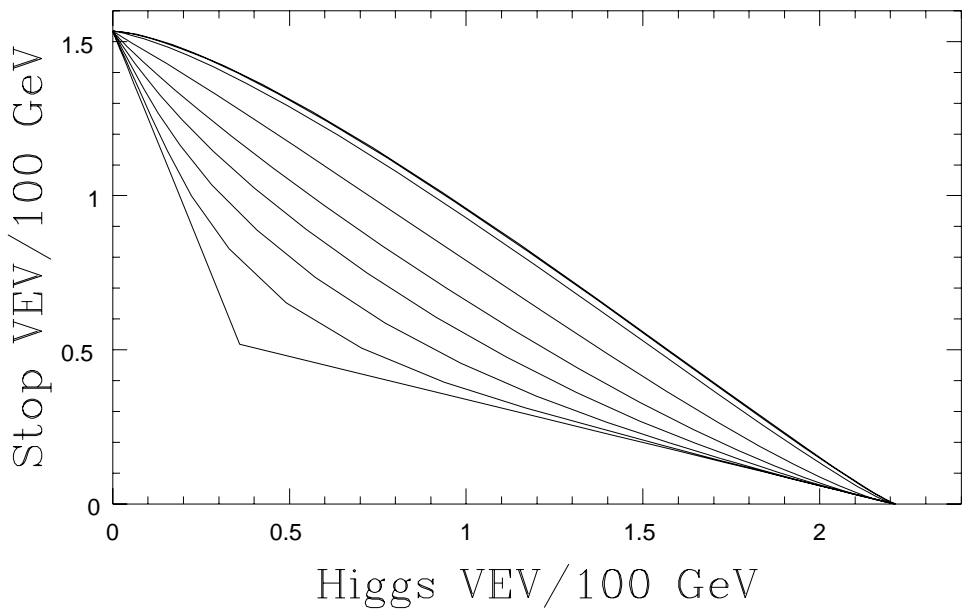}}
\caption{\label{move_through_fieldspace} An example of how the second
algorithm converges to the right line through field space.  The starting
guess for the line through field space is the leftmost one with a sharp
angle, and each line in the series represents the result of one more
iteration of the algorithm.  The algorithm converges quickly to the
right line through field space.}
\end{figure}

Apart from step size errors, the algorithm converges to a saddle point
configuration with only one unstable direction.  This is because the
shooting procedure only allows one unstable mode, associated with
variations in dependence of the fields on the radius while staying on
the same path, and the gradient descent algorithm does not tolerate any
unstable modes for which the fields leave the path.  There is no
guarantee that we will find the lowest action; if there are several
saddlepoint solutions with only one unstable direction, the one we find
depends on the basin of attraction in which the starting guess for a
path lies.  This is a general problem with any saddle point seeking
algorithm.  However we have not found it to be a problem in practice.

We have compared this algorithm with the one described in the last
subsection.  They converge to the same solutions and give the same
saddle point energy to about $1\%$ accuracy for the step size we use.
It is easier to make the general algorithm give higher accuracy; one
recomputes the action with half the step size and extrapolates to zero
step size assuming $O(\Delta^2)$ errors.  This leaves a very small
$O(\Delta^4)$ error which in practice can be made of order $10^{-4}$
quite easily.  We have been less successful bringing the errors of the
algorithm presented here below $O(\Delta^2)$.  However, the algorithm
efficiency is drastically better, especially when the saddle point
action is large; and since we are neglecting corrections (such as
vacuum two-loop contributions to $V$, field dependent wave function
corrections, and higher derivative corrections) which enter at the
$1\%$ level we see little point in pursuing numerical accuracy
further.

\section{Renormalization Group choice of couplings}
\label{AppendixB}

Here we discuss the renormalization group analysis, used to determine the
scalar couplings at a renormalization point $\overline{\mu}$.  To begin
with, we need values for the strong and Yukawa couplings.
We take the value of the strong coupling in the
five quark scheme at the $Z$ pole, $\alpha_s(91{\rm \:
 GeV},\overline{\rm MS})=0.118$, and convert it to $\overline{\rm DR}$ in the
six quark plus right squark scheme using the relation \cite{Pierce}
\beqa
g_s^2(M_Z,\overline{\rm DR},6 {\rm \; quark+right \; squark}) & = &
	\frac{\alpha_s(M_Z,\overline{\rm MS},5 {\rm \; quark})}{1- \Delta
	\alpha_s} \, , \\
\Delta \alpha_s & = & \frac{\alpha_s}{2 \pi} \left[ \frac{1}{2} -
	\frac{2}{3} \ln \frac{m_t}{m_Z} - \frac{1}{6} \ln 
	\frac{m_{s}}{m_Z} \right] \, ,
\eeqa
which coincidentally gives almost the same value.  We run this to the
top mass using the one loop beta function, to be given shortly.
We also determine the
Yukawa coupling at $\overline{\mu}=m_t$ from the expression
\cite{Pierce} 
\beq
\frac{y \sin \beta(\overline{\rm DR},\overline{\mu}=m_t)}{\sqrt{2}} 
	= \frac{m_t}{v} \left( 1 - 
	\frac{5 g_s^2}{12 \pi^2} \right) \, .
\eeq
In $\overline{\rm MS}$ the $5$ would be a $4$.  We define 
$\beta$ so that $\sin
\beta$ is the overlap between the light and up-type ($H_2$) Higgs eigenstates
using the wave functions at the renormalization point set by
the heavy Higgs field threshold; below the threshold only the
combination $y \sin \beta$, which is the coupling of the light Higgs to
the top quark, appears.  The exception is the top-stop-Higgsino
coupling, which we approximate to be $1/\sin \beta$ times the
Higgs-top-top coupling.

We run $g_s^2$ and $y^2$ to the ultraviolet using one loop beta
functions, including only strong and Yukawa contributions in the beta
functions, and putting each heavy particle into loops after crossing its
threshold.  At the energy scale of the heaviest particle, we relate
$\lambda_s$ and $\lambda_h$ to the gauge couplings using the SUSY
relations, given in the main text in 
Eq.\ (\ref{tree_relations}); similarly 
\beq
\lambda_y(\overline{\mu}=UV) = y^2 - \frac{1}{3} g'^2
\eeq
fixes $\lambda_y$ above all thresholds.  These SUSY relations
hold at this UV scale, although if we had used $\overline{\rm MS}$ there
would be nonlogarithmic one loop corrections.  
Then we run all 5 couplings back down to the
infrared, switching to the effective
theory without a heavy particle when we cross its mass threshold.  We
allow ourselves the approximation that the Yukawa-like couplings of
gluinos and Higgsinos equal the respective strong and Yukawa couplings.
Although these relationships are actually broken below heavy particle
thresholds we believe that this produces only a small error.  We also
systematically drop electroweak contributions to the beta functions.

The procedure is possible because the strong and Yukawa beta functions
do not depend on the scalar self-couplings; otherwise we would have to
seek UV values of $g_s^2$ and $y^2$ which would ``hit'' the appropriate
IR values.  The procedure is necessary because our choices for particle
masses lead to large logarithms like $\log(m_Q/m_t) \simeq 4$, which
makes it important to include, for instance, two-loop $\log^2$
contributions.  The difference between performing the renormalization
group analysis and simply enforcing the SUSY relations between the
couplings at our infrared renormalization point is a shift of order $20
\%$ in $\lambda_s$ and $\lambda_y$, and of course a larger shift in
$\lambda_h$, which has a small SUSY value at low $\tan \beta$ but large
radiative corrections from the Yukawa coupling.  The residual two loop
and electroweak errors left out from our analysis 
should be of order a few percent.

Now we present the complete expressions for the beta functions.  The
simplest is the strong beta function,
\beqa
\beta_{g_s^2} & = & \frac{g_s^4}{16 \pi^2} \left\{ -\frac{41}{3}
	+4\; \theta(\overline{\mu}-m_{\tilde{g}})
	+\frac{2}{3}\; \theta(\overline{\mu}-m_Q)
	+3\; \theta(\overline{\mu}-m_{\rm heavy}) \right\} \, .
\eeqa
Here $-41/3$ is the value in the six quark standard model plus right
stop, and the $\theta$
functions turn on each particle's contribution as $\overline{\mu}$
passes its mass threshold; the sum of the terms is $-6$, which is the
correct expression in the full SUSY theory.  

The expressions for the other couplings
are less elegant; for the Yukawa coupling we have
\beqa
\beta_{y^2} & = & \frac{y^2}{16 \pi^2} \Bigg[ 9 y^2 \sin^2 \beta 
	- 16 g_s^2 +9y^2 \cos^2\! \beta\,
	\theta(\overline{\mu}-m_{A^0}) + \nonumber \\ & &
	+\left( 2 y^2 + \frac{8}{3} g_s^2 \right)  
	\theta(\overline{\mu}-m_Q) 
	+y^2 \theta(\overline{\mu}-m_{\tilde{h}})
	+\frac{8}{3}\; g_s^2 \theta(\overline{\mu}-m_{\tilde{g}}) 
	\Bigg] \, ,
\eeqa
where the dependence on $m_{A^0}$ is because we actually change what we
mean when we cross its threshold.  Above the $A^0$ threshold, 
the Yukawa coupling is the
coupling of the up-type ($H_2$) Higgs field to the tops; below, $y^2 \sin^2
\beta$ is the coupling of the light Higgs field to the tops.
The expression below all mass thresholds agrees with the standard model
value and the result above thresholds agrees with the MSSM result.

The expressions for the scalars are even more complicated.  For the
squark self-coupling, and using SUSY relations for its couplings via D
terms to other squarks (which are heavy, so the SUSY relations hold when
it matters), we have
\beqa
\beta_{\lambda_s} & = & \frac{1}{16 \pi^2} \Bigg[ \frac{13}{6} g_s^4
	- 16 g_s^2 \lambda_s + 28 \lambda_s^2 +2 \lambda_y^2
	+\frac{3}{2} g_s^4 \theta(\overline{\mu}-m_{\rm heavy})+
	\nonumber \\ & & 
	+\left(\frac{1}{3} g_s^4 - \frac{4}{3} g_s^2 y^2
	+ 2 y^4\right) \theta(\overline{\mu}-m_Q)
	+\left( \frac{32}{3} g_s^2 \lambda_s - \frac{44}{9} g_s^4
	\right) \theta(\overline{\mu}-m_{\tilde{g}}) \nonumber \\ &&
	+(8 y^2 \lambda_s - 4y^4) 
	\theta(\overline{\mu}-m_{\tilde{h}}) \Bigg]\, ,
\eeqa
where the reader should be cautious because the meaning of $\lambda_y$
in this expression changes at $m_{A^0}$ and $m_Q$; at $m_{A^0}$ it goes
from being the coupling between the up-type Higgs and stop to that of
 the light Higgs and stop, and at $m_Q$ it is modified
by mixing, reducing it by a factor of $1-(\tilde{A}^2/m_Q^2)$.  As
previously noted we assume $m_{A^0}=m_Q$ for simplicity.

To match $\lambda_y$ across the $m_Q$ threshold, we require that
$m_{A^0} = m_Q$.  There are two threshold effects; first, the coupling
of the light Higgs below the threshold is $\sin^2\! \beta$ times the
coupling of the up type Higgs to the stop, plus $\cos^2\! \beta$ times the
coupling of the down type Higgs to the stop, which is $g'^2/3$.  Also,
there is the mixing induced by the diagram in Figure \ref{shift_fig}.
The matching condition across the threshold is therefore
\beq
\lambda_y({\rm below}) = \lambda_y({\rm above}) \sin^2\! \beta
	+ \frac{g'^2}{3} \cos^2\! \beta 
	- y^2 \sin^2\! \beta \frac{\tilde{A}^2}{m_Q^2} \, .
\eeq

The expression for the beta function of $\lambda_y$, valid both above and
below the $m_{A^0}$ threshold, is
\beqa
\beta_{\lambda_y} &=& \frac{1}{16 \pi^2} \Bigg[ 6 \lambda_y y^2 \sin^2\!
	\beta + (4 \lambda_y + 12 \lambda_h + 16 \lambda_s - 8 g_s^2) 
	\lambda_y + 2 y^4 \theta(\overline{\mu}-m_Q)
	+ \nonumber \\ & & 
	\left( 6 y^2 \lambda_y - \frac{32}{3} g_s^2-4 y^4 \right)
	\cos^2\! \beta \theta(\overline{\mu}-m_{A^0})
	+(4 y^2 \lambda_y - 4 y^4 \sin^2\! \beta)
	\theta(\overline{\mu}-m_{\tilde{h}}) \nonumber \\ &&
	+\left( \frac{16}{3} g_s^2 \lambda_y - 
	\frac{32}{3} g_s^2 y^2 \sin^2\! \beta \right)
	\theta(\overline{\mu}-m_{\tilde{g}}) \Bigg] \, .
\eeqa
Including $\lambda_h$ effects in the beta function for $\lambda_y$ is
slightly inconsistent because $\lambda_h$ is largely an electroweak
effect and the canceling electroweak effect required by SUSY is missing
since we ignore electroweak couplings.  However the error this causes is
negligible because $3 \lambda_s \lambda_h/(4 \pi^2)$ is numerically very
small compared to $\lambda_y$.  

Lastly there is the beta function for $\lambda_h$.  It barely runs
above $m_Q$, so we enforce its SUSY relation
there, choosing the value just below to be $(g^2 + g'^2) \cos^2 (2
\beta)/8$.  Below both $m_Q$ and $m_{A^0}$ thresholds, we run it using the
beta function
\beqa
\beta_{\lambda_h} = \frac{1}{16 \pi^2} \Bigg[ 3 \lambda_y^2
	-6 y^4 \sin^4 \beta + 12 y^2 \lambda_h \sin^2 \beta \Bigg] \, .
\eeqa
The electroweak
correction to $\lambda_h$ from the very heavy squarks is not entirely
negligible, because of the large log and because $\lambda_h$ is not very
big; it shifts the final value of
$\lambda_h$ by about $5\%$ of the SUSY value.  We have neglected this
effect in our work, as part of consistently dropping electroweak
radiative corrections, which is reasonable because the Yukawa type
corrections to $\lambda_h$ are of order 1.


\end{document}